\documentclass[useAMS,usenatbib]{mn2e}
\usepackage{color,graphicx} % NB you need to use graphicx or else you can't use include graphics command
\usepackage{latexsym, times, setspace, amssymb, amsfonts, rotating} 
\usepackage{natbib}
\usepackage[fleqn]{amsmath} % make equations flush to the left
\usepackage[integrals]{wasysym}
\usepackage[tight]{subfigure}
\usepackage{siunitx}  % SI units package
\usepackage{lscape}

\begin{document}

\title[Apparent Nulls and Variability in J1107-5907]{On the Apparent Nulls and Extreme Variability of PSR~J1107$-$5907}

\date{\today}

\author[N.~J.~Young et al.]{N.~J.~Young,$^{1,2}$\thanks{E-mail:
    Neil.Young@wits.ac.za} P.~Weltevrede,$^2$ B.~W.~Stappers,$^2$
  A.~G.~Lyne$^2$ and
  M.~Kramer$^{2,3}$\vspace{0.4cm}\\ \parbox{\textwidth}{$^1$School of Physics,
    University of the Witwatersrand, PO BOX Wits, Johannesburg, 2050, South
    Africa \\$^2$Jodrell Bank Centre for Astrophysics, School of Physics, The
    University of Manchester, Manchester M13 9PL, UK
    \\ $^3$Max-Planck-Institut f\"{u}r Radioastronomie, Auf dem H\"{u}gel 69,
    53121 Bonn, Germany}} \maketitle
\begin{abstract}
We present an analysis of the emission behaviour of PSR~J1107$-$5907, a source
known to exhibit separate modes of emission, using observations obtained over
approximately 10~yr. We find that the object exhibits two distinct modes of
emission; a strong mode with a broad profile and a weak mode with a narrow
profile. During the strong mode of emission, the pulsar typically radiates
very energetic emission over sequences of $\sim200-6000$ pulses
($\sim60$~s~$-$~24~min), with apparent nulls over time-scales of up to a few
pulses at a time. Emission during the weak mode is observed outside of these
strong-mode sequences and manifests as occasional bursts of up to a few
clearly detectable pulses at a time, as well as low-level underlying emission
which is only detected through profile integration. This implies that the
previously described null mode may in fact be representative of the bottom-end
of the pulse-intensity distribution for the source. This is supported by the
dramatic pulse-to-pulse intensity modulation and rarity of exceptionally
bright pulses observed during both modes of emission. Coupled with the fact
that the source could be interpreted as a rotating radio transient (RRAT)-like
object for the vast majority of the time, if placed at a further distance, we
advance that this object likely represents a bridge between RRATs and extreme
moding pulsars. Further to these emission properties, we also show that the
source is consistent with being a near-aligned rotator and that it does not
exhibit any measurable spin-down rate variation. These results suggest that
nulls observed in other intermittent objects may in fact be representative of
very weak emission without the need for complete cessation. As such, we argue
that longer ($\gtrsim1$~h) observations of pulsars are required to discern
their true modulation properties.

\end{abstract}

\section{Introduction}\label{sec:intro}
PSR~J1107$-$5907 is an old, isolated radio pulsar which was discovered in the
Parkes 20-cm Multibeam Pulsar Survey of the Galactic plane \citep{lfl+06}. It
has a rotational period ($P\sim253$~ms) which is normal among the pulsar
population. However, its period derivative ($\dot{P}\sim9\times10^{-18}$) is
comparatively low, thus placing the object in an underpopulated region in
\hbox{$P-\dot{P}$} space; that is, between the population of normal and
recycled pulsars, which is home to only a small percentage of the total
population. As a further consequence, the inferred characteristic age of the
source ($\tau_{\mathrm{c}}\sim447$~Myr) also indicates that it is amongst the
oldest $\sim4$~per~cent of known non-recycled pulsars\footnote{See
  http://www.atnf.csiro.au/people/pulsar/psrcat/ for published data on
  currently known sources.}.

In addition to these interesting characteristics, a study by \cite{okl+06}
indicated that the neutron star alternates between a null (or radio-off) state
where no emission is detectable, a weak mode which has a narrow profile and a
bright mode which exhibits a very broad profile (see left panel of
Fig.~\ref{profiles}). In the bright emission state, during their observations,
the object was often observed to saturate their one-bit filterbank system and,
subsequently, was deemed to be of comparable brightness to the Vela pulsar
(see also \citealt{obr10})\footnote{PSR~B0833$-$45, a.k.a. the Vela pulsar,
  emits single pulses with peak flux densities ranging up to $\sim10^2$~Jy at
  1410~MHz (e.g. \citealt{kjv02}).}. Due to the low cadence of their
observations, the mode-switching time-scales associated with the source could
not be firmly constrained. Instead, it was shown that the source could cycle
between its separate emission modes over long time-scales ($\sim$~h), which
are considered to be atypical of `normal' nulling pulsars ($\lesssim100\,P$;
e.g.  \citealt{wmj07}). Furthermore, a connection with the longer-term
intermittent pulsar B1931$+$24 \citep{klo+06,ysl+13} was noted, due to the
long time the source appeared in its null state.

In the only other published study of the source, \cite{bjb+12} discovered an
isolated single pulse from the object in one of their HTRU med-lat survey
observations. Combined with the discovery of more regular emission in archival
Parkes observations, they inferred that the source exhibits different nulling
fractions (NFs) in each active emission mode, similar to that observed in
PSRs~B0826$-$34 and J0941$-$39 \citep{bjb+12}. Remarkably, both of these
objects appear to switch between an emission mode with similar properties to
rotating radio transients (RRATs; \citealt{mll+06})~$-$~with single pulse
detection rates of $\sim100$~h$^{-1}$ and $\sim50$~h$^{-1}$
respectively~$-$~and a more typical `pulsar-like' emission mode where the
objects are detected more regularly \citep{bb10,bjb+12,eamn12}. PSR~B0826$-$34
has also been shown to exhibit very weak emission, which can be confused with
apparent null phases without sufficient pulse integration \citep{elg+05};
c.f. null confusion in PSRs J1648$-$4458 and J1658$-$4306 \citep{wmj07}. This
behaviour has been likened to the evolutionary progression of a pulsar towards
its `death', where it no longer emits radio emission
\citep{zgd07,bb10}. However, no firm connection has been made to date.

Several theories have been proposed to explain the moding and/or transient
behaviour of RRATs and other pulsars in the above context; e.g. temporary
reactivation or enhancement of emission due to the presence of circumstellar
asteroids \citep{cs08}, magnetic field instabilities \citep{grg03,ug04,wmj07}
and surface temperature variations in the polar gap region
\citep{zqlh97}. Each of these trigger mechanisms can be consolidated with a
scenario where the pulsar undergoes rapid changes in its magnetospheric charge
distribution (e.g. \citealt{tim10,lhk+10,lst12a,hhk+13}). However, none is
able to fully describe how such changes, or degradation, in the radio emission
mechanism could occur. This is further compounded by the lack of a fully
self-consistent model of how radio emission is produced in the pulsar
magnetosphere (see e.g. \citealt{kkhc12}).

Since the initial analysis by \cite{okl+06}, ongoing observations of this
source have been made using the Parkes 64-m telescope. With the increase in
the number of observations, and availability of single-pulse data, a more
detailed study of the emission and rotational characteristics of
PSR~J1107$-$5907 has been made possible, which is presented in this paper. We
will subsequently show that the source only exhibits two modes of
emission~$-$~a strong mode and a weak mode~$-$~during which very weak emission
can be confused with nulls, analogous to that seen in PSR~B0826$-$34 and a
handful of other pulsars. Coupled with the fact that the source is one of only
a few known objects to exhibit intermediate moding time-scales
(i.e. $\sim$~min to hr; see \citealt{kle+10}), PSR~J1107$-$5907 thus
represents an ideal target for studying the potential range of emission
variability in pulsars. In the following section we describe the observations
of PSR~J1107$-$5907. This is followed by an overview of its emission
properties in Section~\ref{sec:em_props} and timing analysis in
Section~\ref{sec:timing}. Lastly, we discuss the implications of our results
in Section~\ref{sec:discuss}, in the context of other pulsars and emission
modulation theories, and summarize our conclusions in Section~\ref{sec:conc}.

\vspace{-5.6mm}
\section{Observations}\label{sec:obs}
Our data set comprises observations taken from three observing programmes, all
of which were carried out using the Parkes~\hbox{64-m} radio telescope. These
observing programmes made use of the H-OH, Multibeam and 1050cm receivers,
each of which has dual-orthogonal linear feeds (see e.g. \citealt{mhb+13} for
detailed specifications.).

The majority of the data used in this paper come from an intermittent source
monitoring programme (IMP), carried out between 2003~February~21 and
2010~August~24, using the H-OH receiver and central beam of the Multibeam
receiver (refer to Table~\ref{tab:obs}). These observations were recorded
using an analogue filterbank system which one-bit digitized the data at
$987~\mu$s intervals. These data were later folded off-line at the pulsar
period to produce both folded data with sub-integration intervals of
$\sim59.3$~s and single-pulse archives\footnote{Single-pulse data were not
  obtained for 10 of the archival observations.}.

The second portion of our data comes from a dedicated set of multi-frequency
observations taken in the period 2012~October~18-20, using the central beam of
the Multibeam receiver and the dual frequency 1050cm receiver (refer to
Table~\ref{tab:select_obs}). Two separate digital filterbanks were used to
record the 20-cm data, at $256~\mu$s and 60~s (sub-integration) intervals
using 8-bit digitization. Single-pulse observations were formed through
off-line folding. A polarized calibration signal was also injected into the
receiver probes, and observed, prior to each of the 20-cm observations in
order to polarization calibrate the data. By comparison, the 10- and 50-cm
observations were obtained simultaneously in one observing session on
2012~October~20, using one digital filterbank per frequency band (see
Table~\ref{tab:select_obs}). These data were sampled at $256~\mu$s intervals
using 8-bit digitization, and were later folded offline to form single-pulse
archives and 60~s sub-integration data.

The last portion of our data used in this paper was obtained through a recent
intermittent source monitoring programme (IMP2; 2011~May~2 to
2012~November~22), with sole use of the central beam of the Multibeam receiver
(refer to Table~\ref{tab:obs}). A digital filterbank system was used to record
these data at $494~\mu$s intervals using 8-bit digitization. The observations
were also folded off-line to form 10-s sub-integration data.

\begin{table*}
\caption{The observation characteristics of the intermittent monitoring
  programmes (IMP). OBSREF denotes the reference given to the extended
  observing programmes, MJD represents the modified Julian Date at the start
  of the observations and $T_{\mathrm{span}}$ refers to the total time-span of
  the observations. The total number of observations carried out is given by
  $N_{\mathrm{obs}}$, $T_{\mathrm{obs}}$ denotes the typical observation
  duration and $\langle C\rangle$ represents the average observation
  cadence. The centre sky frequency, observation bandwidth and total number of
  frequency channels are denoted by $\nu$, $\Delta \nu$ and
  $N_{\mathrm{chan}}$, respectively.}  \centering
\begin{tabular}{ c c c c c c c c c c }
  \hline
   OBSREF & MJD & $T_{\mathrm{span}}$ (d) & $N_{\mathrm{obs}}$ & $T_{\mathrm{obs}}$ (min) & $\langle C\rangle$ (d$^{-1}$) &  $\nu$ (MHz) & $\Delta \nu$ (MHz) & $N_{\mathrm{chan}}$\\
  \hline
  IMP-Multi & 52691.7 & 2741 & 274 & 5   & 0.10 & 1374 & 288  & 96   \\
  IMP-HOH   & 52984.0 & 1244 & 65  & 15  & 0.05 & 1518 & 576  & 192  \\
  IMP2      & 55683.4 & 571  & 37  & 10  & 0.07 & 1369 & 256  & 1024 \\
  \hline
\end{tabular}
\label{tab:obs}
\end{table*}

In off-line processing, we de-dispersed and examined the data for
radio-frequency interference (RFI). As emission is observable throughout most
of the pulse period, and because of the variability in the source's
brightness, automated RFI mitigation through conventional signal thresholding
methods was infeasible. Therefore, frequency channels and single
pulses/sub-integrations particularly affected by RFI were manually flagged and
weighted to zero with {\small\textsc{PSRZAP}} and
{\small\textsc{PAZ}}\footnote{See \cite{hvm04} and
  http://psrchive.sourceforge.net/manuals for details on these software
  packages.}. Further RFI analysis was also performed on the pulse intensity
distributions of the longest observations. Here, a custom-made script was used
to compute the skewness, kurtosis, variance, total intensity and most extreme
negative values of each pulse. Those pulses which exhibited one or more of
these quantities above a certain threshold were flagged and analysed by eye,
before also being weighted to zero.  While the overall data quality is quite
good, we find that a number of archival observations are badly affected by RFI
and/or are subject to saturation due to particularly bright emission from the
source. Therefore, these observations are only used to help infer the
time-scales of emission variation.

We estimated flux values through measurement of the peak signal-to-noise ratio
(SNR) of each profile and inserted these values into the single-pulse or
modified radiometer equation, depending on the number of pulses integrated
over (see e.g. \citealt{mc03,lk05}), along with the known observing system
parameters. This method of flux calibration results in typical errors of
$\sim30$~per~cent (see e.g. \citealt{kle+10}).

\begin{table*}
\caption{The properties of the dedicated multi-frequency observations.  The
  reference key for each observation is denoted by REF and is in the format of
  `YYMMDD'$-$`observing band'. MJD is the modified Julian date at the start of
  each observation, $N$ is the total number of single-pulses and
  $N_{\mathrm{zap}}$ is the number of pulses weighted to zero. The total
  observation length is denoted by $T$ and `mode' represents which emission
  states are observed.}  \centering
\begin{tabular}{ c c c c c c c c c }
  \hline
  REF & MJD & $\nu$ (MHz) & $\Delta \nu$ (MHz) & $N$ & $N_{\mathrm{zap}}$ & $T$ (s) & Mode\\\hline
  121018$-$20cm & 56218.8986 & 1369 & 256  & 85277 & 13070  & 21556 & Strong+weak \\
  121019$-$20cm & 56219.7656 & 1369 & 256  & 70875 & 11034  & 17915 & Strong+weak \\
  121020$-$10cm & 56220.0316 & 3094 & 1024 & 48325 & 8491   & 12215 & Weak \\
  121020$-$50cm & 56220.0316 & 732  & 64   & 48326 & 13204  & 12215 & Weak \\
  \hline
\end{tabular}
\label{tab:select_obs}
\end{table*}

\vspace{-3.8mm}
\section{Emission Properties}\label{sec:em_props}
Previous works have shown that PSR~J1107$-$5907 is a highly variable source,
which exhibits peculiar moding behaviour and apparent nulls
\citep{okl+06,obr10,bjb+12}. However, the time-scales of these variations have
previously not been constrained. Nor has there been an in-depth investigation
into the salient characteristics attributed to the source or its apparent
nulling activity. To rectify this, we present a detailed review of the
emission properties of PSR~J1107$-$5907, and describe how they can be used to
differentiate between the separate emission modes in the following
subsections.

\subsection{Variability time-scales}\label{sec:mod}
We initially set out to constrain the time-scales of emission variation in
PSR~J1107$-$5907 by analysing the average profiles of each observation. Of the
380 total observations analysed, we find that 210 show detectable radio
emission. Among these detections, the pulsar is found to exhibit its strong
emission mode in 22 observations ($\sim6$~per~cent of the total). In 12 of
these strong-mode dominated observations, the pulsar also displays weak-mode
emission. The remaining 188 detections show the source solely in its weak
emission state ($\sim50$~per~cent of the total).

\begin{figure*}
\begin{center}
 % NB Trim dimensions are left,bottom,right,top 
  \includegraphics[trim = 20mm 0mm 25mm 20mm, clip,height=6.5cm,width=8.4cm]{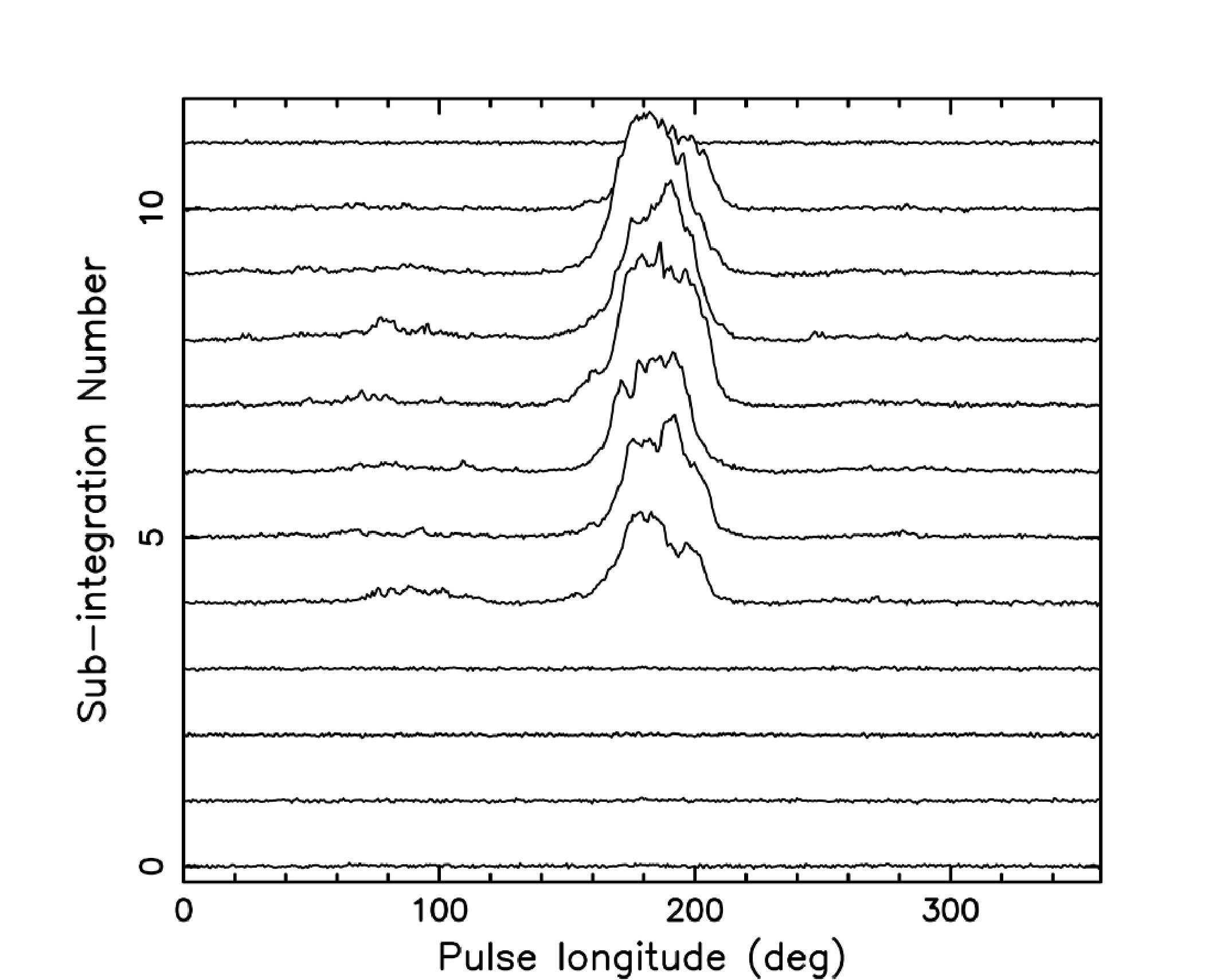}
 % NB Trim dimensions are left,bottom,right,top 
  \includegraphics[trim = 20mm 0mm 25mm 20mm, clip,height=6.5cm,width=8.4cm]{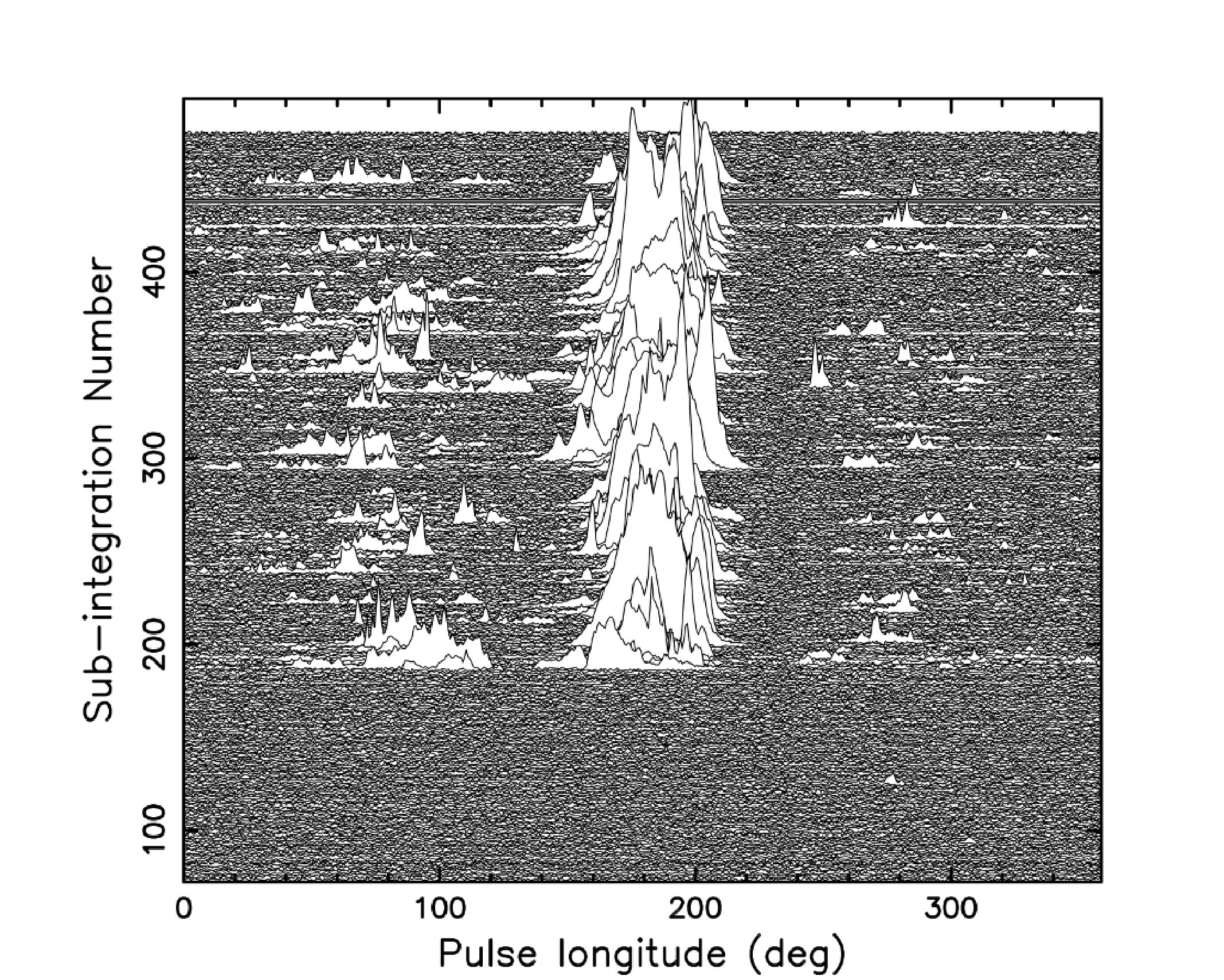}
 % NB Trim dimensions are left,bottom,right,top 
  \includegraphics[trim = 20mm 0mm 25mm 20mm, clip,height=6.5cm,width=8.4cm]{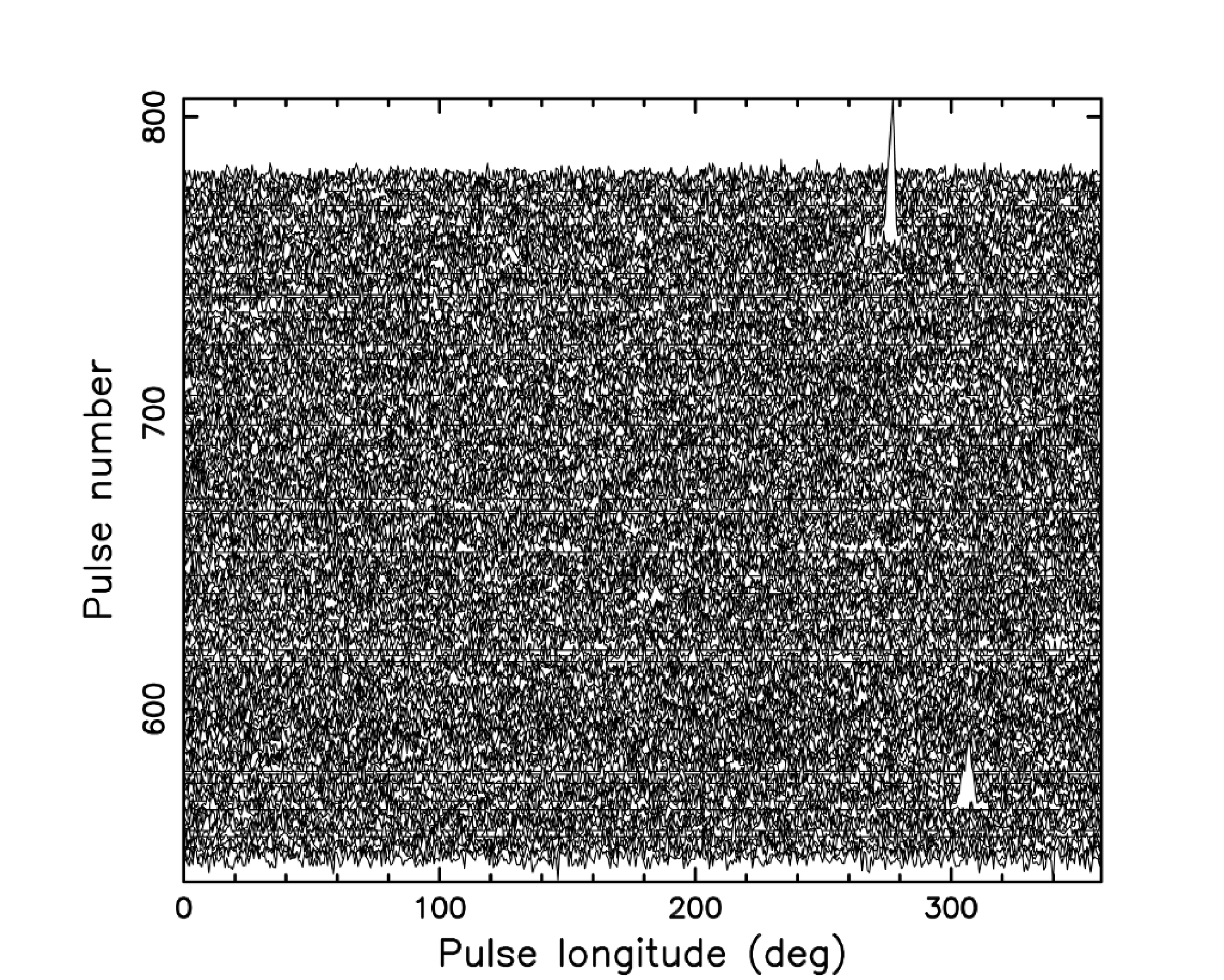}
 % NB Trim dimensions are left,bottom,right,top 
  \includegraphics[trim = 20mm 0mm 25mm 20mm, clip,height=6.5cm,width=8.4cm]{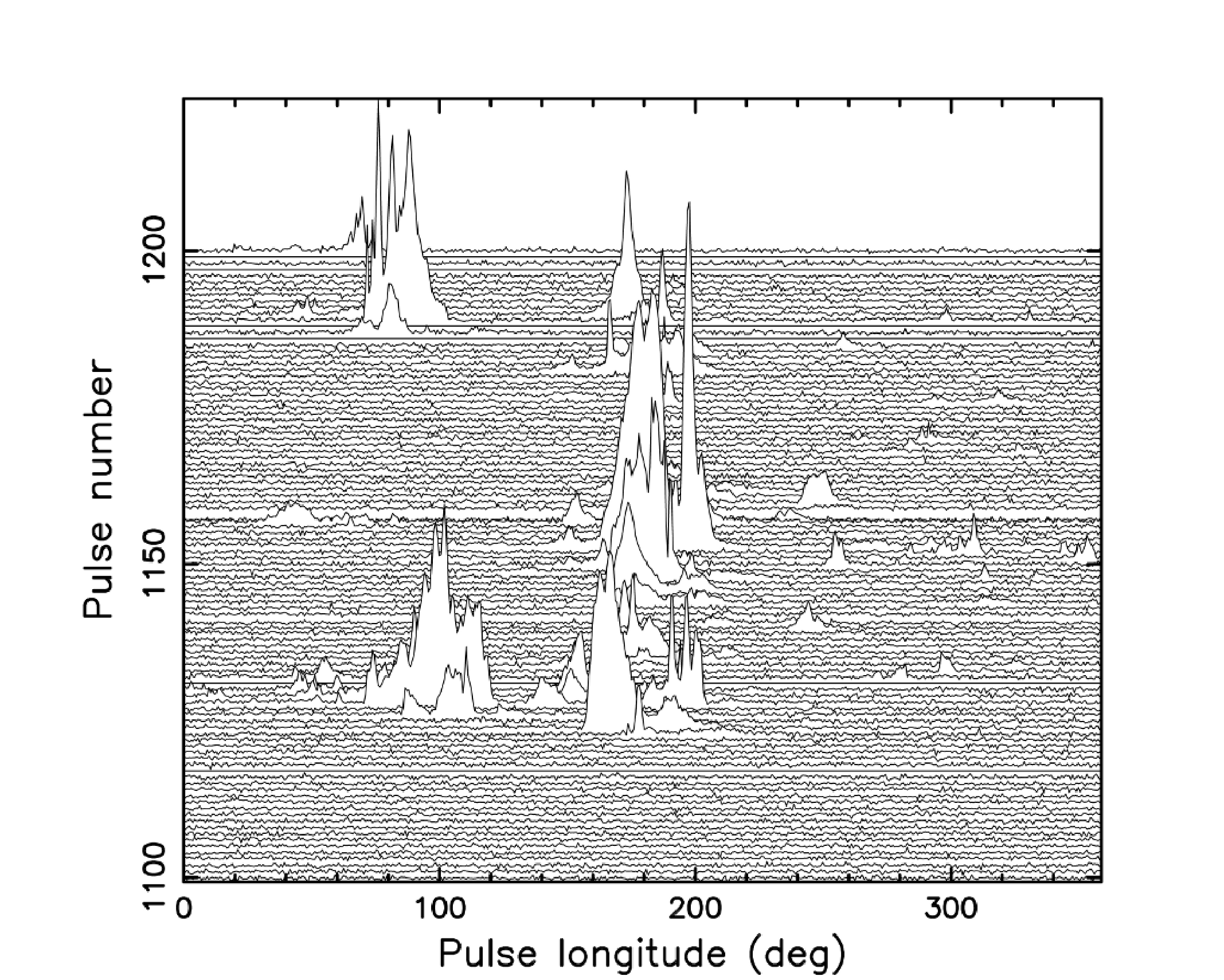}
\end{center}
\vspace{-5pt} 
\caption{\emph{Top left:} integrated pulse profiles formed from $\sim63$~s
  sub-integrations (i.e. 250 pulses) for an $\sim13$~min extract (i.e. 3000
  pulses) from 121018$-$20cm. The sub-integration data is separated into
  consecutive profiles along the vertical axis (from bottom to top). The
  pulsar appears to only become active (and bright) in the fifth
  sub-integration. \emph{Top right:} pulse stack of a slightly shorter data
  extract (pulses $450-2850$), plotted using sub-integration lengths of six
  pulses only ($\sim1.5$~s). The greater time resolution afforded by these
  data display the short-time-scale modulation of the source which is not
  resolved when averaging over many pulses. \emph{Bottom left:} single-pulse
  data for pulses $550-780$, showing two clearly detected weak-pulses during
  the otherwise apparent null phase. \emph{Bottom right:} single-pulse data
  for pulses $1100-1200$, showing how the pulsar transitions between apparent
  null phases and the strong-mode of emission.}
\label{transitions}
\end{figure*}

While, the above statistics offer an interesting insight into the moding
behaviour of the source, they do not present the whole picture attributed to
the source's variability, particularly at short time-scales. Concentrating our
focus on the longest \hbox{($\gtrsim3$~h)} high quality observations, for
which single-pulse data were available (see Table~\ref{tab:select_obs}), we in
fact find that the source behaves more like a highly variable moding pulsar
than a source which undergoes longer stable radio-on and -off phases
(c.f. PSR~B1931$+$24; \citealt{klo+06,ysl+13}). Fig.~\ref{transitions}
demonstrates this pulse-to-pulse variability for the observation 121018$-$20cm
(refer to Table~\ref{tab:select_obs}).

We find that this short-time-scale variability is inconsistent with
interstellar scintillation, given the object's narrow scintillation bandwidth
\hbox{($\lesssim40$~MHz)} and relatively long scintillation time-scales at
1518~MHz ($\sim~5$~min, predicted by the {\small\textsc{NE2001}} model;
\citealt{cl02}). As a result, we are confident that the emission modulation
observed in this object is intrinsic.

\vspace{-4.5mm}
\subsection{Mode Description and Durations}\label{sec:modes}
Due to the pulse-to-pulse variability of the source, emission profiles formed
from sub-integration data, i.e. the average of $\sim10^{2-3}$ pulses, will not
provide an accurate representation of the source's emission characteristics or
variability. Therefore, we used a boxcar algorithm with a variable width to
locate single pulses with significant peaks, and facilitate correct
characterization of the pulse properties attributed to each emission
mode. Using our highest quality observations (see Table~\ref{tab:select_obs}),
we find that a $6\sigma$ sensitivity limit results in the most reliable
location of discernible single-pulse emission\footnote{As the source can emit
  over almost its entire pulse window, only narrow OP regions can be
  defined. Subsequently, the root-mean-square variation and average amplitude
  attributed to the noise of each pulse cannot be accurately determined. This,
  in turn, can lead to spurious SNR measurements and more frequent
  false-positive detections for lower significance limits.}. From those
observations which were long enough to provide sufficient coverage of the
source's emission behaviour (i.e. the 20-cm data), we find that density of
peak detections is directly related to the mode in which the pulsar
assumes. Namely, during the weak mode of emission, peak detections arise
sporadically between apparent null phases up to several hundred pulse
periods. By contrast, peak detections are grouped in dense clusters of pulses
with apparent nulls up to a few pulse periods only during the strong mode.

\begin{figure*}
\begin{center}
 % NB Trim dimensions are left,bottom,right,top 
  \includegraphics[trim = 20mm 0mm 25mm 20mm, clip,height=5.5cm,width=8.4cm]{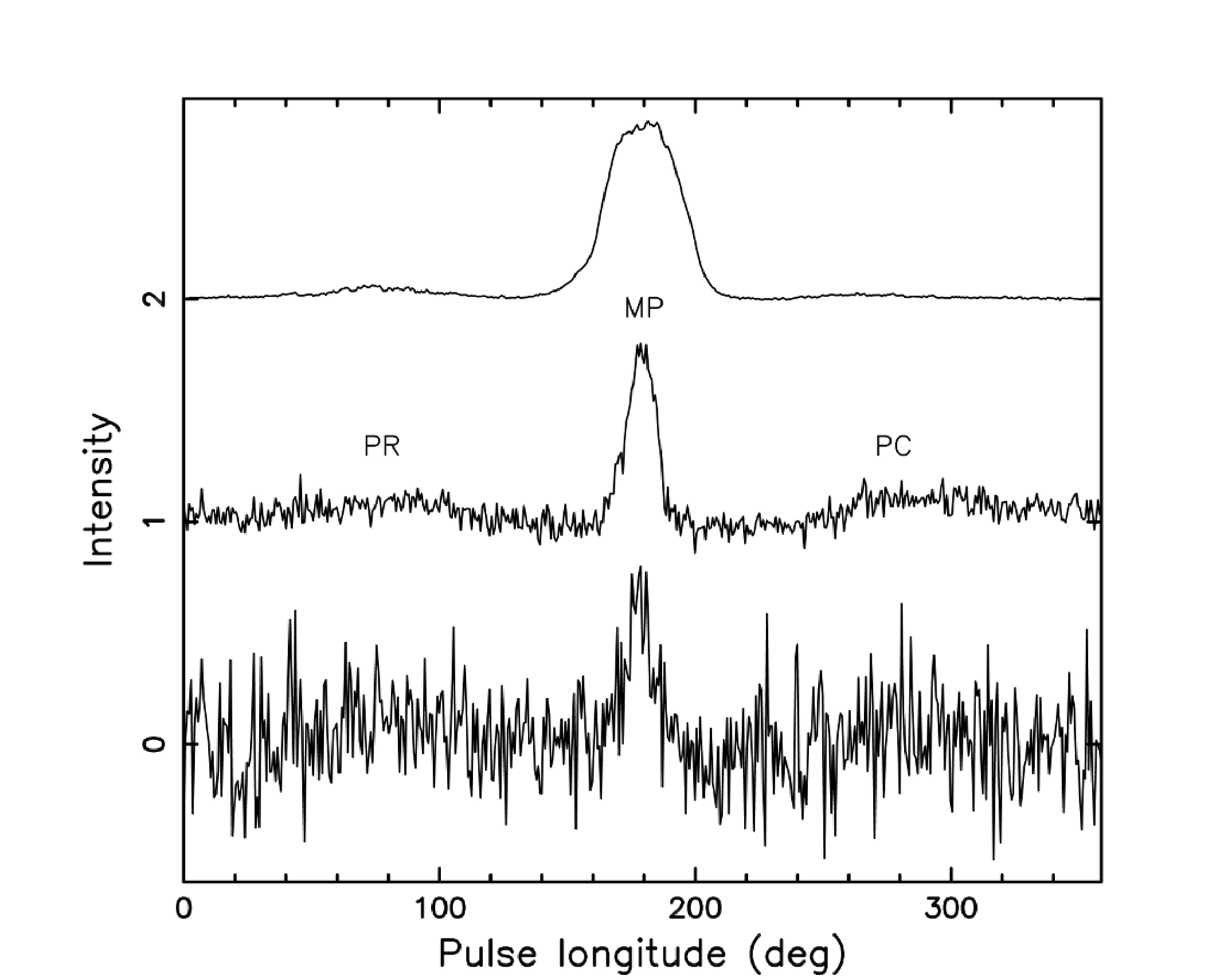}
 % NB Trim dimensions are left,bottom,right,top 
  \includegraphics[trim = 14mm 12mm 25mm 17mm, clip,height=5.5cm,width=8.4cm]{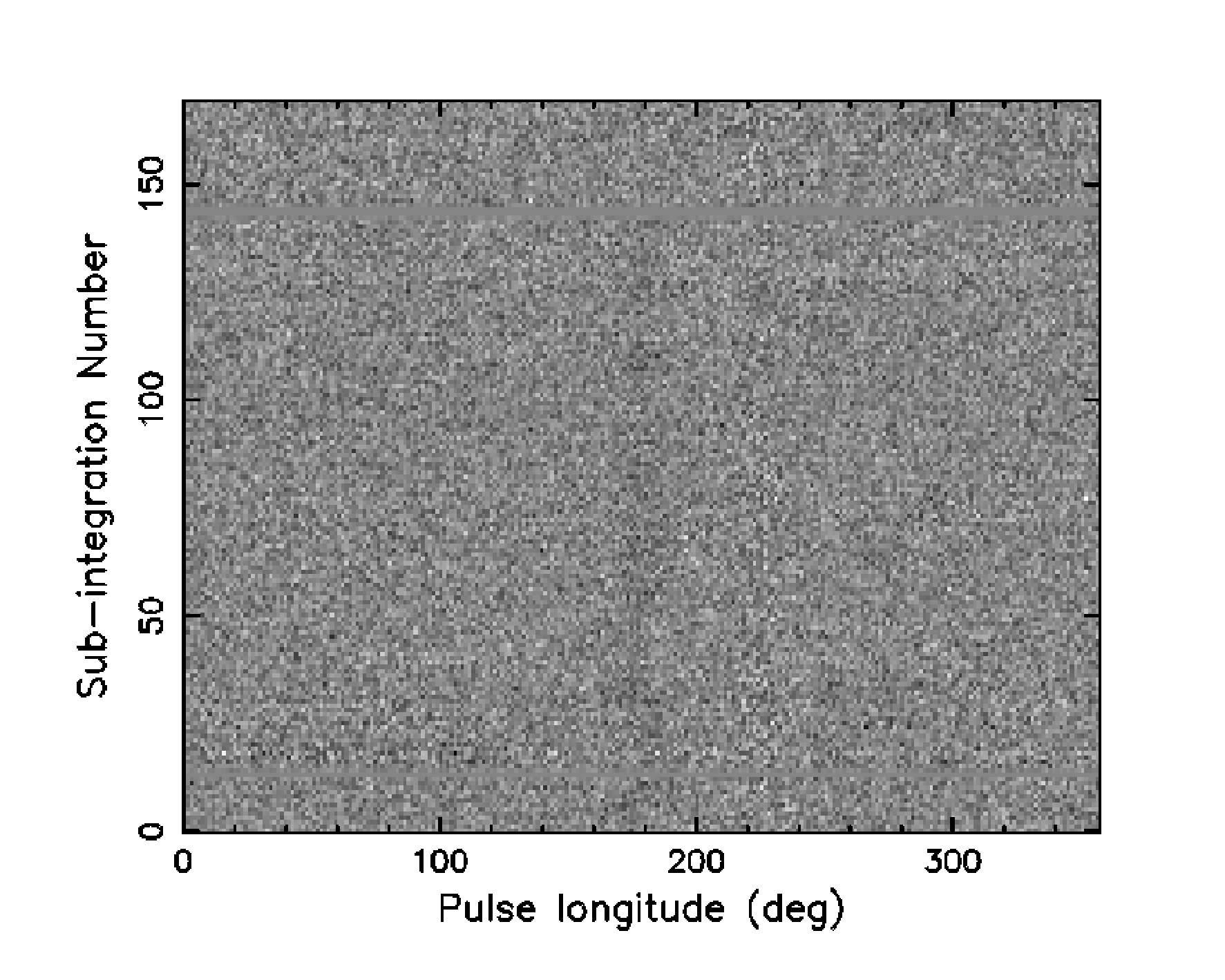}
\end{center}
\vspace{-7pt}
\caption{\emph{Left panel:} from top to bottom, the average profiles formed
  from 121018$-$20~cm are shown for the strong and weak emission modes as well
  as for the UE during the apparent null phase. The on-pulse emission regions
  are also annotated. Each of the integrated profiles contains 1452, 70755 and
  8360 pulses, respectively. Note that the pulse profiles are vertically
  offset from each other for clarity, and are normalized by their respective
  peak intensities. \emph{Right panel:} grey-scale intensity map showing how
  the UE during the apparent null phase varies with time, after rebinning the
  data to 256 bins and integrating over consecutive groups of 500 pulses. Note
  that while the MP emission is clearly visible, the outer profile components
  are still below the noise threshold of the data.}
\label{profiles}
\end{figure*}

During the aptly named strong mode, we find that emission from the source is
significantly enhanced, and hence more readily detectable, compared with that
of the sporadically detected weak pulses. Furthermore, the pulsar exhibits a
considerably broader main-pulse (MP) component in the average pulse profile
than that in the weak mode. Interestingly, our data shows that strong-mode
pulses are only emitted during relatively short burst periods ($\sim60$~s up
to $\sim24$~min), with an average duration of approximately 500~s and a
standard deviation of $\sim400$~s (see Figs.~\ref{transitions} and
$-$\ref{profiles}). It is important to note, however, that 11 out of the 18
total detected bursts are not completely covered by our
observations. Therefore, the quoted average duration for the emission bursts
serves as a lower bound to the true value.

Upon closer inspection of the apparent null pulses, which are far more
prevalent during the weak mode, we in fact find that the vast majority of them
contain emission which is comparable to or below the detection thresholds of
our data. This becomes clear upon averaging subsequent sequences of pulses
($\sim10^2-10^3$~pulses), where low-level or underlying emission (UE) becomes
more readily detectable with the number of pulses which are integrated over
(see right panel of Fig.~\ref{profiles}). Thus, we use the term apparent null
to define pulses which are very weak and can be easily confused with null
emission (c.f. \citealt{elg+05,wmj07}).

Due to the prevalence of apparent nulls in the pulsar, average profiles were
formed in a few ways. That is, we formed average profiles with respect to the
separate emission modes, and observing frequency, for all pulses and for only
the $6\sigma$ detected pulses in the highest quality observations (see
Table~\ref{tab:select_obs}). For comparison, we also produced low-level
emission profiles for the same observations, by locating and averaging pulses
with no detectable peaks. From these data it is clear that while the average
profiles of both the strong and weak modes exhibit emission both prior to and
after the MP component~$-$~which we refer to as precursor (PR) and postcursor
(PC) emission components, respectively\footnote{\cite{rib08} refers to the PR
  and PC emission as a single, broad inter-pulse component.}~$-$~these
components typically constitute a more significant proportion of the average
emission profile during the weak mode, compared with that of the strong
mode. We also find that the source does not exhibit a stable profile over the
time-scales of our observations, in either the strong or weak modes of
emission, regardless of the number of pulses integrated over
(c.f. PSR~B0656$+$14; \citealt{wwsr06}). This is particularly evident during
the strong mode of emission, which exhibits significant profile variations
between successive observations.

Furthermore, we note that the pulses containing UE, during the apparent null
state, are uniformly distributed throughout the data, which suggests that the
pulsar may not truly undergo any conventional null phases (see e.g.
Fig.~\ref{profiles}). Rather, clearly detected weak pulses could represent
those which are at the top end of the source's pulse energy distribution (PED)
during the otherwise apparent null mode. As such, we advance that the apparent
null phases most likely do not represent a discrete emission state, and that
they only constitute the lowest end of the PEDs of the strong and weak
emission modes.

In order to further constrain the pulse properties of the separate emission
modes, we estimated the mean flux densities attributed to the pulsar, using
the method outlined in Section~\ref{sec:obs}, from the mode-separated average
profiles. We also searched the high quality observations for the maximum
peak-flux pulse densities to provide a direct brightness comparison with
RRATs. Table~\ref{tab:flux} shows the results of this
analysis.\nocite{mc03,lk05,cl02,mhb+13}

\begin{table*}
\caption{The observed emission properties of PSR~J1107$-$5907. The columns are
  as follows: (1)~type of emission from the pulsar (UE = underlying emission).
  (2)~Centre frequency of the observation.  (3)~Equivalent pulse width of the
  brightest pulse which results in a boxcar with the same peak and integrated
  intensities as the observed pulse ($\omega=P\times\Sigma
  I/I_{\mathrm{peak}}$). (4)~Peak flux density of the brightest pulse
  determined by using the single-pulse radiometer equation (see e.g.
  McLaughlin~$\&$~Cordes 2003) and the receiver system specifications quoted
  (Manchester et al. 2013). (5)~Peak pseudo-luminosity of the pulsar estimated
  by the equation $L_{\mathrm{peak}} = S_{\mathrm{peak}}\,d^2$
  (Lorimer~$\&$~Kramer 2005), where $d=1.28$~kpc is the distance to the source
  from the {\small\textsc{NE2001}} model (Cordes~$\&$~Lazio 2002). (6)~Number
  of pulses used to compute the average emission properties.  (7)~Equivalent
  pulse width attributed to the average emission profile. (8)~Mean flux
  density of the pulsar, calculated using the modified radiometer equation for
  the average profile (see e.g. Lorimer~$\&$~Kramer 2005). (9)~Number of
  pulses used to compute the average emission properties for the $6\sigma$
  detected pulses.  (10)~Equivalent pulse width attributed to the average
  emission profile for the $6\sigma$ detected pulses. (11)~Mean flux density
  of the pulsar, calculated using the modified radiometer equation for the
  average profile formed from the $6\sigma$ detected pulses.}  \centering
\begin{tabular}{ c c c c c c c c c c c }
  \hline
  Mode & $\nu$ (MHz) & $\omega$ (ms) & $\mathrm{S}_{\mathrm{peak}}$ (mJy) & $\mathrm{L}_{\mathrm{peak}}$ (Jy~kpc$^2$) & $N$ & $\langle \omega\rangle$ (ms) & $\langle S\rangle$ (mJy) & $N_{6\sigma}$ & $\langle \omega\rangle_{6\sigma}$ (ms) & $\langle S\rangle_{6\sigma}$ (mJy)\\\hline
  Strong & 1369 & 1.00 & 9883 & 16.19 & 1676   & 24.67 & 11.05 & 919  & 24.68 & 14.65 \\
  Weak   & 1369 & 1.74 & 4178 & 6.85  & 130372 & 10.06 & 0.063 & 4056 & 10.05 & 0.55 \\
  UE     & 1369 & $-$  & $-$  & $-$   & 15835  & 9.33  & 0.025 & $-$  & $-$   & $-$  \\\rule{0pt}{2.5ex}
  Weak   & 732  & 4.64 & 5638 & 9.24  & 35122  & 10.19 & 0.090 & 526  & 11.39 & 0.88 \\
  UE     & 732  & $-$  & $-$  & $-$   & 4532   & 4.40  & 0.078 & $-$  & $-$   & $-$  \\\rule{0pt}{2.5ex}
  Weak   & 3094 & 1.54 & 2183 & 3.58  & 39834  & 11.17 & 0.039 & 1206 & 11.90 & 0.25 \\
  UE     & 3094 & $-$  & $-$  & $-$   & 4068   & 4.50  & 0.014 & $-$  & $-$   & $-$  \\
  \hline
\end{tabular}
\label{tab:flux}
\end{table*}

We note that the peak flux densities of the brightest pulses detected during
the strong and weak emission states are quite comparable at 20-cm. This is,
however, in contrast to the large (a factor of $\sim175$) difference in the
flux densities of the mode-separated profiles, which implies that particularly
energetic weak-mode pulses are exceptionally rare (c.f. the strong to UE flux
density ratio of $\sim440$:1). This is indeed observed in the data, given the
large difference between the average flux densities of all mode-separated
pulses and those attributed to the $6\sigma$ detected mode-separated pulses
only. Moreover, we find that weak-mode pulses with flux densities $\geq1$~Jy
only constitute $\sim0.03$~per~cent of the pulses in the 20-cm band
(c.f. $\sim16$~per~cent for the strong mode).

\subsection{Pulse-energy Distributions}\label{sec:PEDs}
\begin{figure}
\begin{center}
  % NB Trim dimensions are top, left, bottom, right
  \includegraphics[trim = 0.5mm 1mm 1mm 3mm, clip,height=4.15cm,width=4.15cm,angle=270]{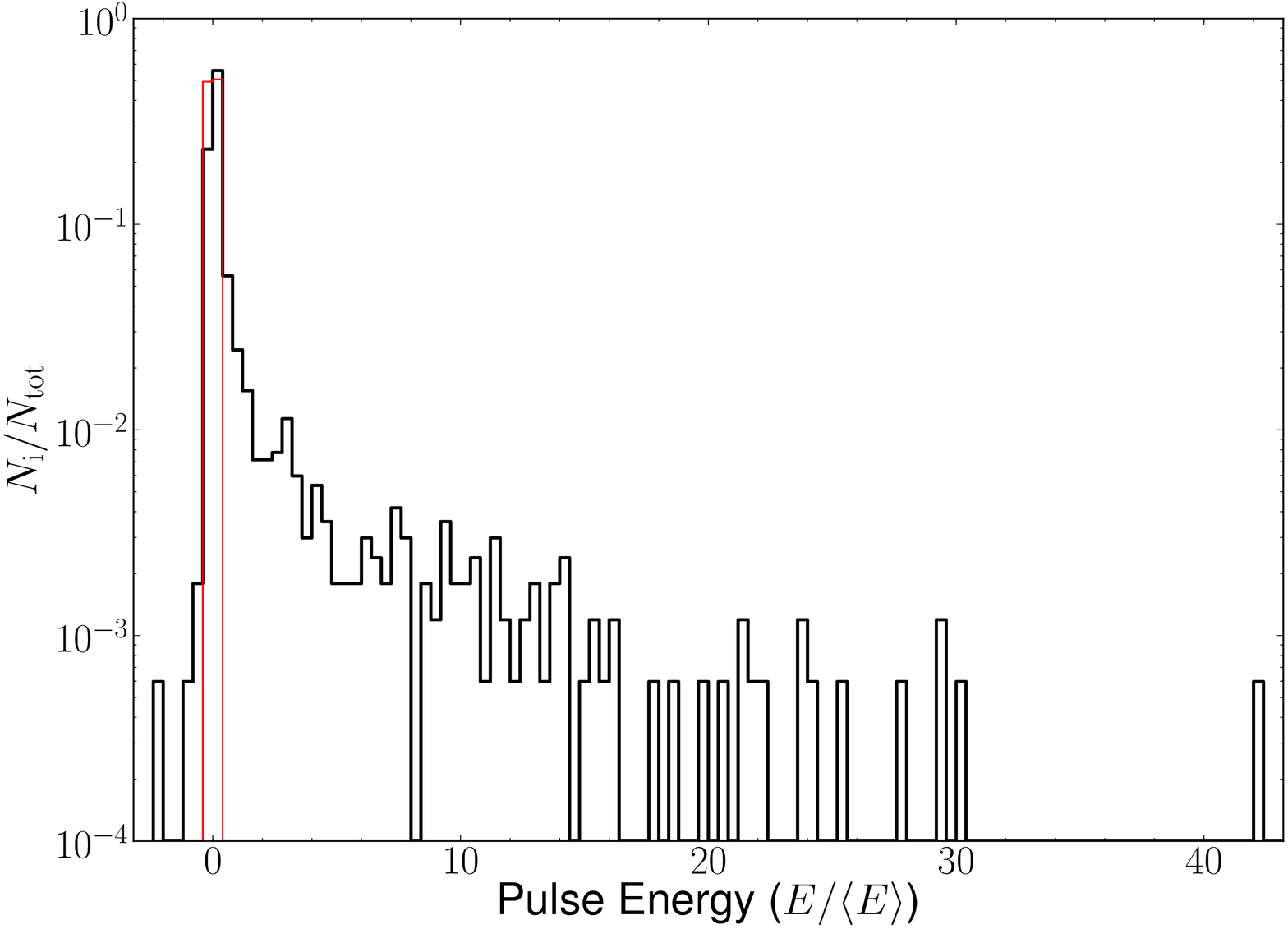}
  \includegraphics[trim = 0.5mm 1mm 1mm 3mm, clip,height=4.15cm,width=4.15cm,angle=270]{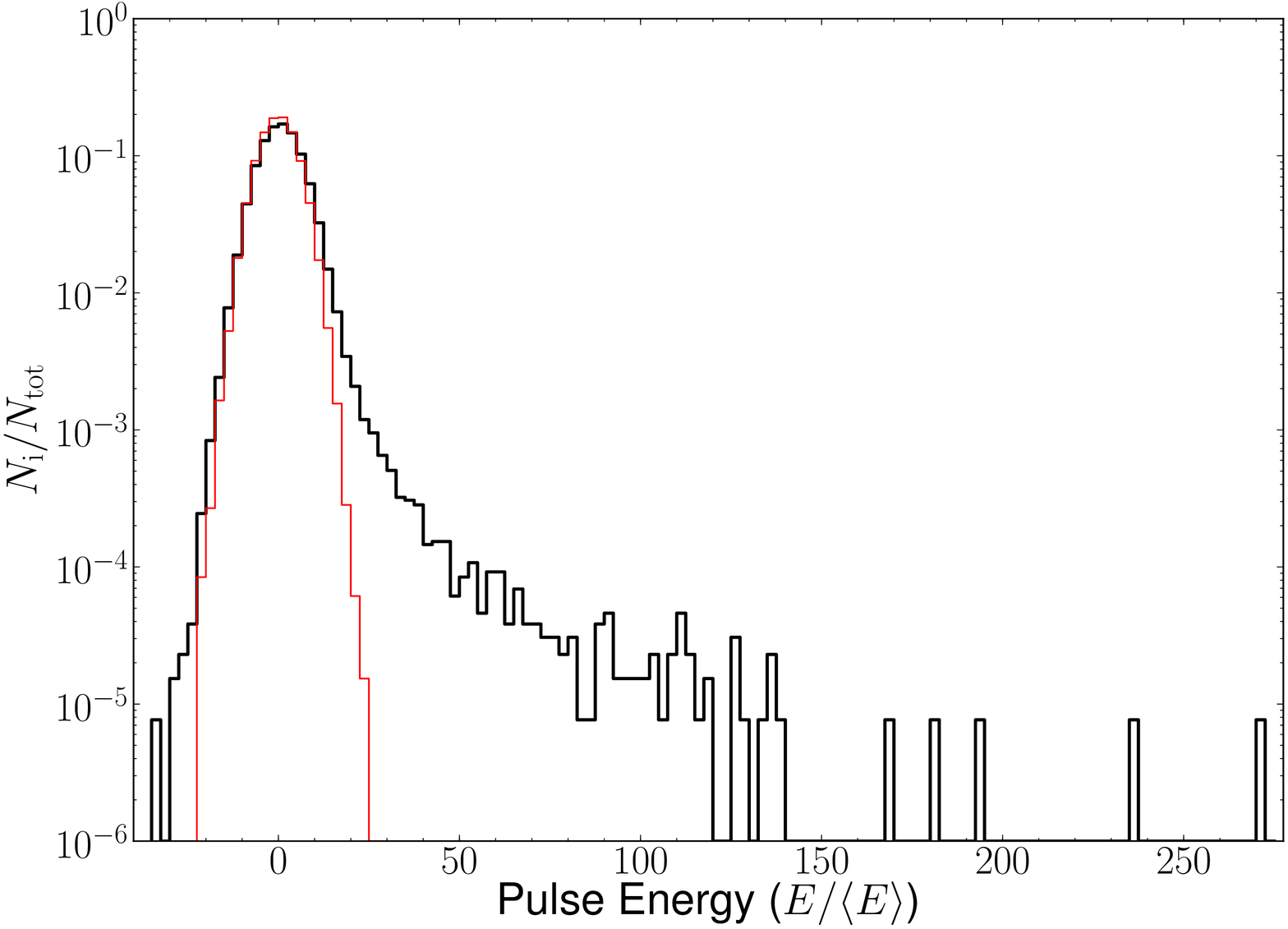}
  \includegraphics[trim = 0.5mm 1mm 1mm 3mm, clip,height=4.15cm,width=4.15cm,angle=270]{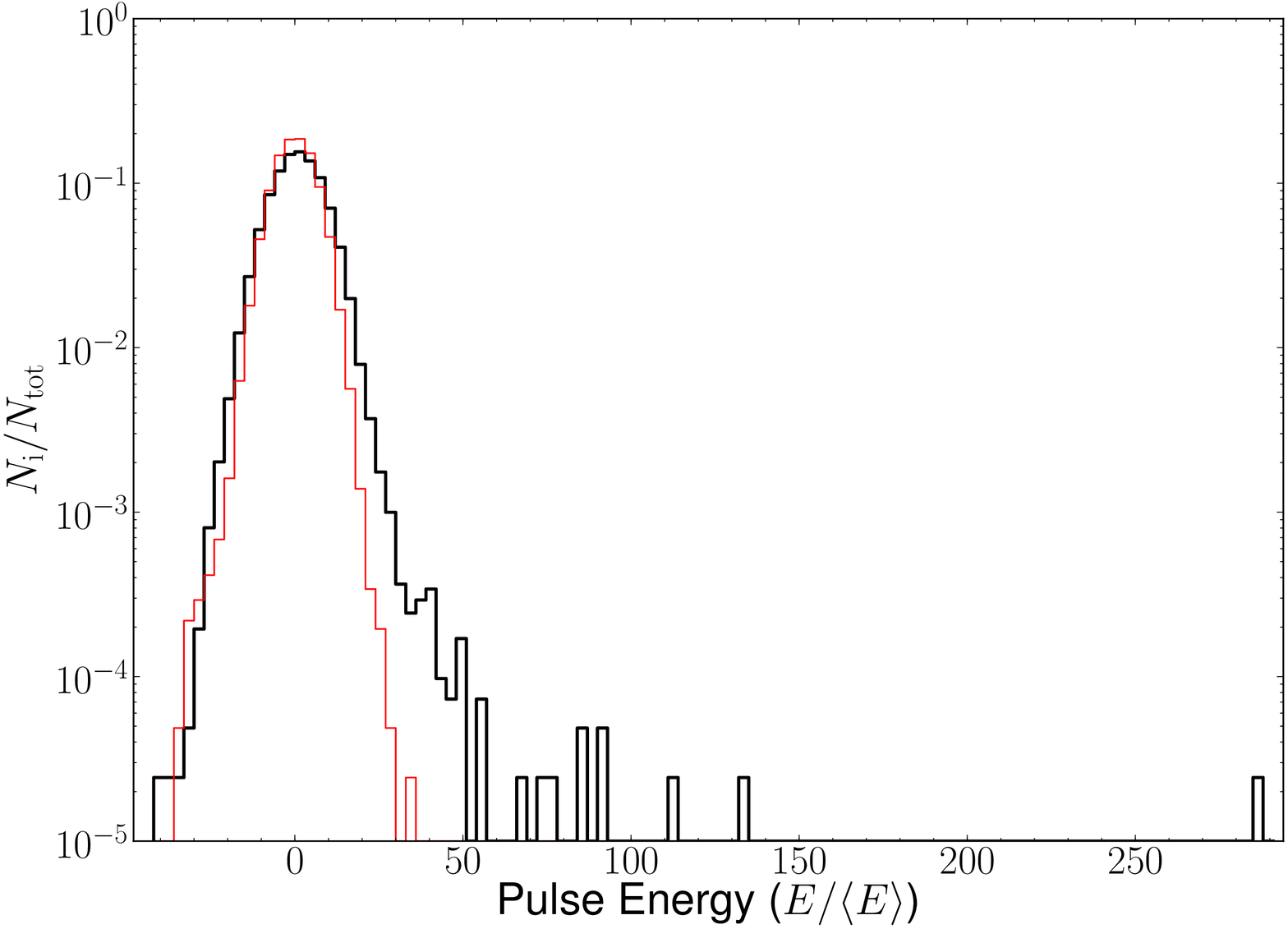}
  \includegraphics[trim = 0.5mm 1mm 1mm 3mm, clip,height=4.15cm,width=4.15cm,angle=270]{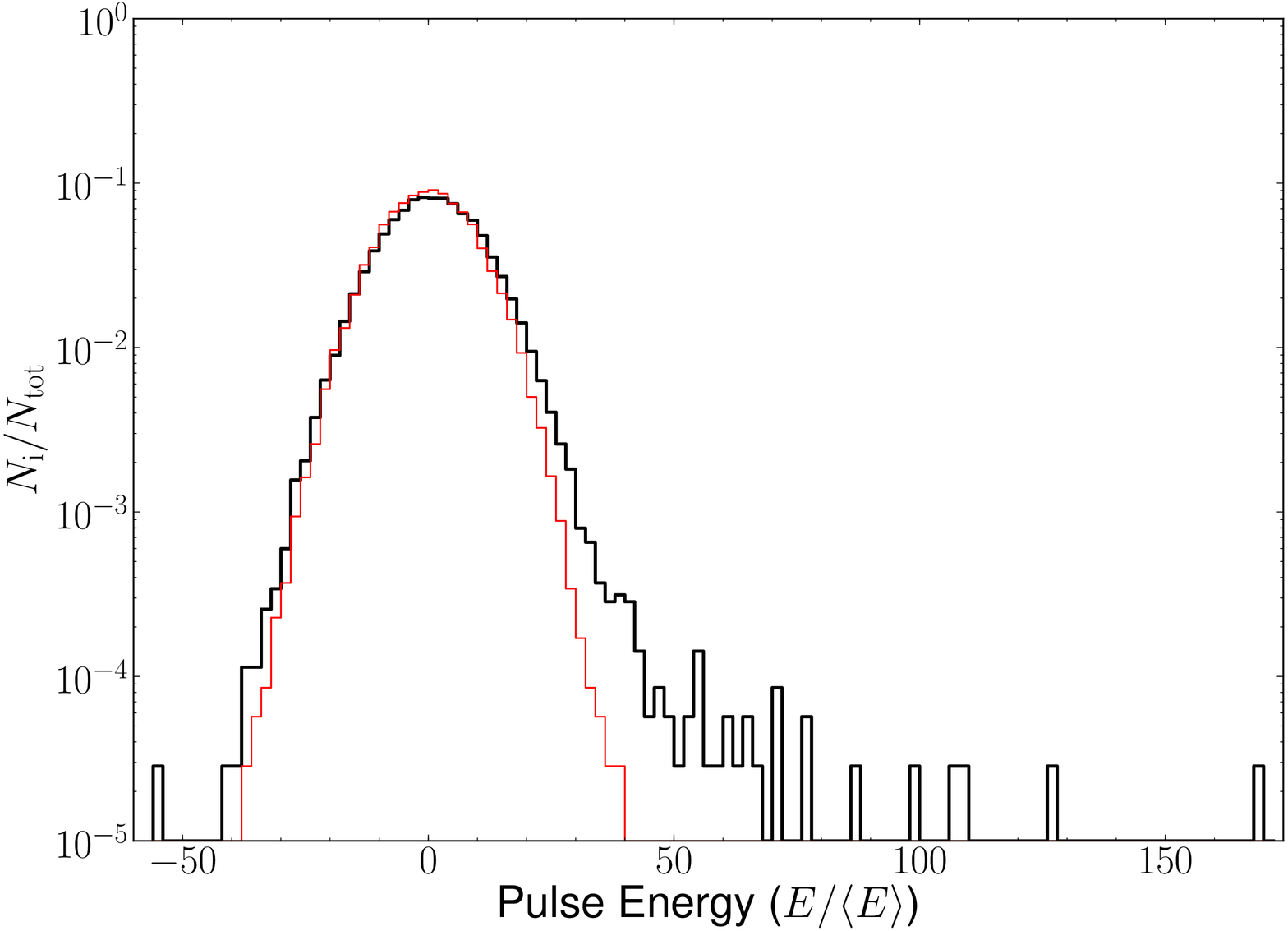}
\end{center}
\vspace{-6.5pt}
\caption{Integrated PEDs for the strong- and weak-mode pulses from the 20-cm
  data (\emph{top, left to right panels, respectively}), as well as those for
  the 10-cm and 50-cm weak-mode pulses (\emph{bottom, left to right panels
    respectively}). The PEDs are normalized by the average pulse energy
  $\langle E\rangle$ of each mode. The MP and OP energies are denoted by the
  \emph{thick black} and \emph{thin red} lines, respectively. The OP energies
  are scaled according to the proportion of MP to OP longitude bins used.}
\label{IPEDs}
\end{figure}

Given the remarkable variability of the source, we sought to further
characterize its intensity fluctuations and apparent nulling behaviour through
computing pulse-energy\footnote{Following \cite{wwsr06}, for instance, we
  define the `pulse energy' to be interchangeable with pulse intensity.}
distributions (PEDs; see \citealt{wwsr06} for details on the method used) for
the mode-separated observations in Table~\ref{tab:select_obs}. For this
analysis, PEDs were formed for the PR, MP and PC regions separately, as well
as for the off-pulse (OP) region and the whole on-pulse regions for each
observation. Due to the difference in the number of pulse longitude bins used,
the OP energies were scaled such that they reflect the predicted noise
fluctuations in the on-pulse regions. We note that no correction for
interstellar scintillation was carried out as the dominant intensity
fluctuations are caused by apparent nulling and mode-changing. Example results
from this analysis are shown in Fig.~\ref{IPEDs}, for the PEDs formed from the
MP region.

For the majority of our data, we find that the PEDs of the PR, PC and whole
on-pulse regions typically do not have as clearly pronounced emission tails as
those of the MP PEDs. This is particularly clear in the strong emission state,
where the MP PED displays a narrower peak at zero PE and a more distinct
emission tail towards large PEs compared with the other PEDs. This is in
contrast to the results obtained in the 50-cm band, where the whole on-pulse
PED shows greater evidence for emission compared with PEDs formed from PEs in
the other profile regions. This disparity in the dominant PEDs can be
explained by the frequency-dependent prevalence of bright emission in the
respective emission windows (see $\S$~\ref{sec:LRPEDs}).

Overall, we do not find any conclusive evidence for nulls in the PEDs of
either the strong or weak emission modes. That is, the PEDs formed from our
data do not exhibit discernible breaks representative of a null distribution
or divergence from a single distribution function. Rather, they appear to be
continuous which is consistent with the hypothesis that the pulsar typically
emits pulses under a relatively low PE regime, with some sporadic energetic
emission that represents the top-end of its PED (particularly in the strong
emission mode).

\subsubsection{Pulse-energy Distribution Fitting}\label{sec:fits}
Typically, the PEDs of pulsars can be represented by single-component
distributions (see e.g. \citealt{ag72,cai04}). However, in the presence of
nulls, a given PED should either be bimodal or exhibit evidence for a
functional transition. Such features in a PED can exist below the noise
level. Therefore, we sought to confirm the results of the above analysis by
fitting either a power-law or log-normal trial distribution to the
mode-separated PEDs, in combination with a distribution of apparent nulls
below the noise level, following the method described in \cite{wwsr06}. The
functional forms of the fitted distributions are:
\begin{eqnarray}
P_{\mathrm{pow}} (E) &\propto& E^{\,p}\,,\label{eq:powerlaw}\\[2mm]
P_{\mathrm{logn}} (E) &=&\frac{\langle E\rangle}{\sqrt{2\pi}\sigma E}\exp
\left[-\left(\ln\frac{E}{\langle E\rangle}-\mu\right)^2/\left(2\sigma^2\right)\right].\label{eq:lognorm}
\end{eqnarray}

Since the power-law distribution extends to infinity, we incorporated a
minimum pulse energy cut-off $E_{\mathrm{min}}$ into the power-law
fit. Therefore, there are two fit parameters for both model distributions;
i.e. $p$ and $E_{\mathrm{min}}$ for the power-law distribution and $\mu$ and
$\sigma$ for the lognormal model distribution.

The effect of noise was accounted for by convolving the noise signature with
the model PED for each observation. We estimated the noise signature by
producing a symmetric distribution from the negative on-pulse energies. This
`mirrored' distribution is probably an oversimplification of the true noise
variation, but provides a more realistic representation of the noise signature
compared to the distribution derived from the very narrow OP region.

The requirement to include very low-level emission (apparent nulls) in a fit
was also tested by optionally adding pulses with zero energy to a model
distribution (before convolving this distribution with that of the noise),
until the average energy $\langle E\rangle$ matched the observed value for a
given set of fit parameters. The optimization was performed by minimizing the
$\chi^2$ between the model and observed distributions (see \citealt{wwsr06}
for details). Error bars on the fit parameters were also determined by finding
the possible range of fit parameters which could still result in acceptable
fits; i.e., with a significance probability above
$5$~per~cent. Table~\ref{tab:fits} shows the result of this analysis.

\begin{table*}
\caption{The best-fit parameters of the model distributions. The reference key
  of each observation is denoted by REF and the type of pulses included in
  each fit is represented by `mode'. The power-law exponent and energy cut-off
  are represented by $p$ and $E_{\mathrm{min}}$, while the log-normal mean and
  standard deviation are given by $\mu$ and $\sigma$ respectively. The
  apparent NF, total $\chi^2$, number of degrees of freedom
  $N_{\mathrm{d.o.f.}}$ and significance probability $P(\chi^2)$ are also
  tabulated.}
\label{tab:fits}
\begin{center}
\begin{tabular}{ c c c c c c c c c c }
  \hline
  REF & Mode & $p$ & $E_{\mathrm{min}}$ & $\mu$ & $\sigma$ & NF (per cent) & $\chi^2$ & $N_{\mathrm{d.o.f.}}$ & $P(\chi^2\mathrm{,}$~per~cent)\\\hline\rule{0pt}{2.5ex}
  121018$-$20cm & Strong & $-1.29^{+0.03}_{-0.02}$ & $400^{+300}_{-100}$ & $-$ & $-$ & $48^{+5}_{-4}$ & 10.5 & 9 & 30.2 \\[0.5ex]
  121019$-$20cm & Strong$^a$ & $-$ & $-$ & $-$ & $-$ & $-$ & $-$ & $-$ & $-$ \\\rule{0pt}{2.5ex}
  121018$-$20cm & Weak & $-$ & $-$ & $5.8^{+0.1}_{-0.1}$  & $0.9^{+0.1}_{-0.1}$ & $91.5^{+0.5}_{-0.5}$ & 56.1 & 22 & 19.2 \\[0.5ex]
  121019$-$20cm & Weak & $-$ & $-$ & $6.1^{+0.1}_{-0.1}$  & $0.9^{+0.1}_{-0.1}$ & $88.9^{+0.3}_{-0.3}$ & 21.5 & 14 & 29.7 \\[0.5ex]
  121020$-$10cm & Weak & $-$ & $-$ & $7.7^{+0.1}_{-0.2}$  & $0.7^{+0.1}_{-0.6}$ & $89^{+1}_{-5}$ & 16.5 & 20 & 97.1 \\[0.5ex]
  121020$-$50cm & Weak & $-$ & $-$ & $8.7^{+0.2}_{-0.3}$  & $0.8^{+0.2}_{-0.5}$ & $93^{+1}_{-4}$ & 32.2 & 24 & 61.7 \\[0.5ex]
  \hline
\end{tabular}
\end{center}
\begin{flushleft}
\hspace{10mm}$^{a}$Insufficient number of pulses to perform the analysis.
\end{flushleft}
\end{table*}

Overall, we find that the best fits are typically obtained for the MP PEDs
during both the strong and weak emission modes (except for the 50-cm data
where the whole on-pulse PEDs provide the best results). We also note that the
strong-mode PEDs are best fit using a power-law distribution as opposed to a
lognormal distribution for the weak mode PEDs. For both modes of emission, an
additional, apparent null distribution is required to converge the
fits. Therefore, a simple, single-component distribution of the chosen
functional forms cannot be used to describe the PEDs of
PSR~J1107$-$5907. However, this does not necessarily provide evidence for the
existence (or absence) of actual emission cessation in the source, as the PED
fitting cannot distinguish between zero PEs and a PED of very weak
UE. Therefore, we advance that the PEDs of the source cannot be purely
described by lognormal/ power-law statistics, and that the UE may possess a
different functional form whose transition is below the noise level.

\subsubsection{Longitude-resolved Pulse-energy Distributions}\label{sec:LRPEDs}
\begin{figure}
\begin{center}
  % NB Trim dimensions are top, left, bottom, right
  \includegraphics[trim = -1mm 0mm -1mm 0mm, clip,height=8.3cm,width=6.2cm,angle=270]{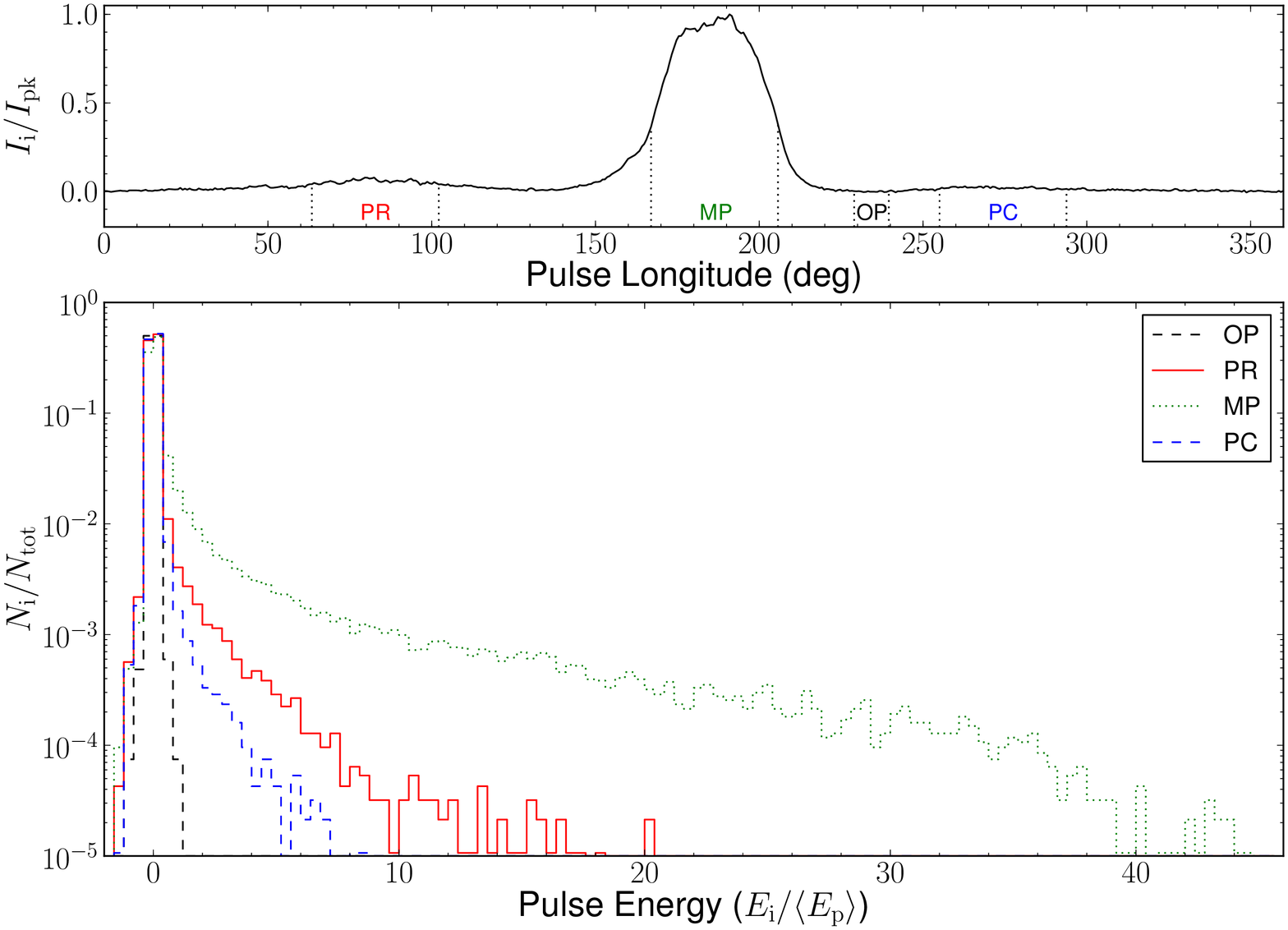}
  \includegraphics[trim = -1mm 0mm -1mm 0mm, clip,height=8.3cm,width=6.2cm,angle=270]{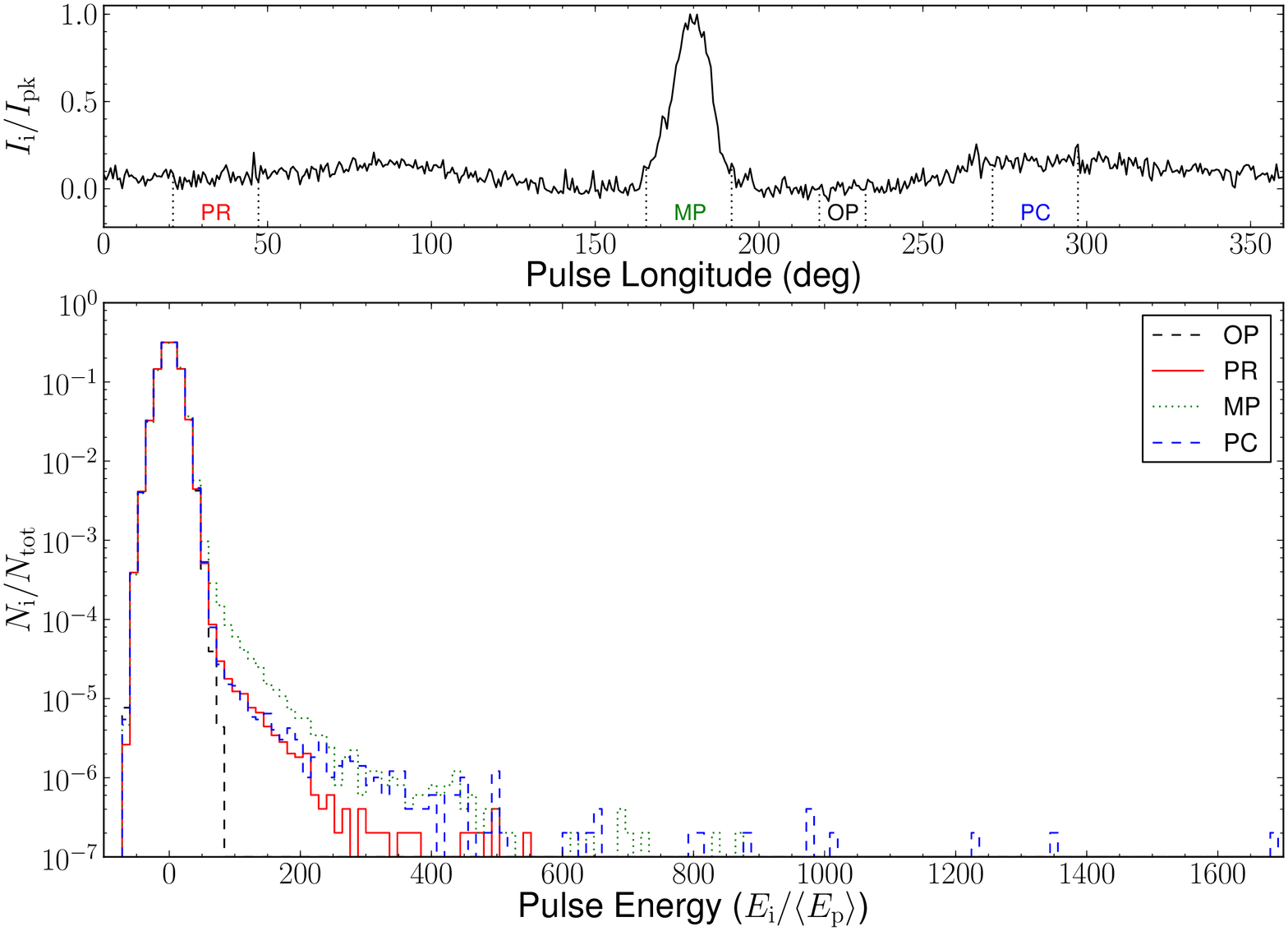}
\end{center}
\vspace{-8pt}
\caption{Longitude-resolved PEDs for the emission-mode separated
  observations. The plots show the results for the strong- and weak-mode
  pulses taken from the combined 20-cm data set (\emph{top and bottom panels,
    respectively}). The top panel for each of these plots shows the average
  profile, normalized by the peak pulse intensity $I_{\mathrm{pk}}$ for each
  observation, with the emission regions annotated. The pulse energies are
  normalized by the average peak-energy $\langle E_{\mathrm{p}} \rangle$.}
\label{LRPEDs}
\end{figure}

While the analysis of the integrated on-pulse energies allows for the
determination of the apparent NF and general pulse properties of the source,
it does not provide a complete picture of the pulse intensity fluctuations
attributed to the object. This is emphasized by the fact that pulsed emission
from pulsars generally exhibits variability as a function of pulse
longitude. Therefore, we sought to characterize such intensity fluctuations,
and determine the dominant emission regions, through computing
longitude-resolved PEDs for the highest quality observations (refer to
Table~\ref{tab:select_obs}). Here, we separated pulse profiles out into four
pulse longitude regions~$-$ the OP, MP and most energetic portions of the PR
and PC emission windows~$-$~and accumulated individual pulse-longitude bin
samples over these regions into separate PEDs, respectively. Fig.~\ref{LRPEDs}
shows the longitude-resolved PEDs which result from this analysis, along with
the corresponding average profiles for the mode-separated observations chosen.

This analysis clearly confirms the prominence of emission in the MP and PR
regions during the strong mode, where we observe a much greater number of high
PE samples compared with that in the PC region. By comparison, emission in the
weak mode is preferentially located to the PC and MP regions. In fact, we see
that the pulsar emits an increasingly higher fraction of high-energy samples
in the PC region during the weak emission mode towards lower frequencies. With
the above in mind, it is clear that the pulse energy characteristics of the
strong and weak modes are quite different which, subsequently, provides
further evidence for a separation in the pulse populations.

\subsubsection{Intensity-dependent Profile Variations}\label{sec:prof_var}
\begin{figure}
\begin{center}
  % NB Trim dimensions are top, left, bottom, right
  \includegraphics[trim = 1mm 1mm 1mm 3mm, clip,height=4.15cm,width=4.15cm,angle=270]{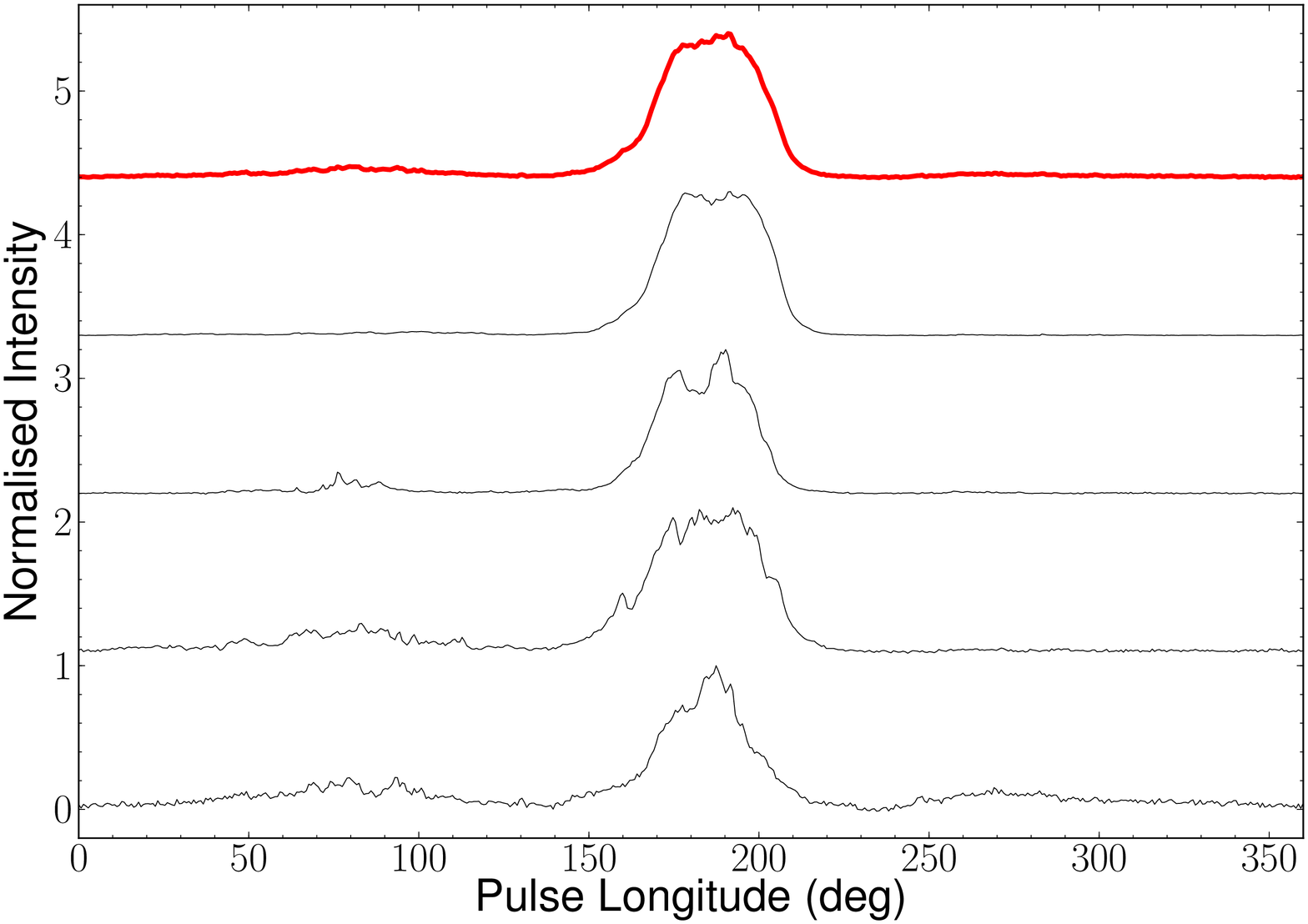}
  \includegraphics[trim = 1mm 1mm 1mm 3mm, clip,height=4.15cm,width=4.15cm,angle=270]{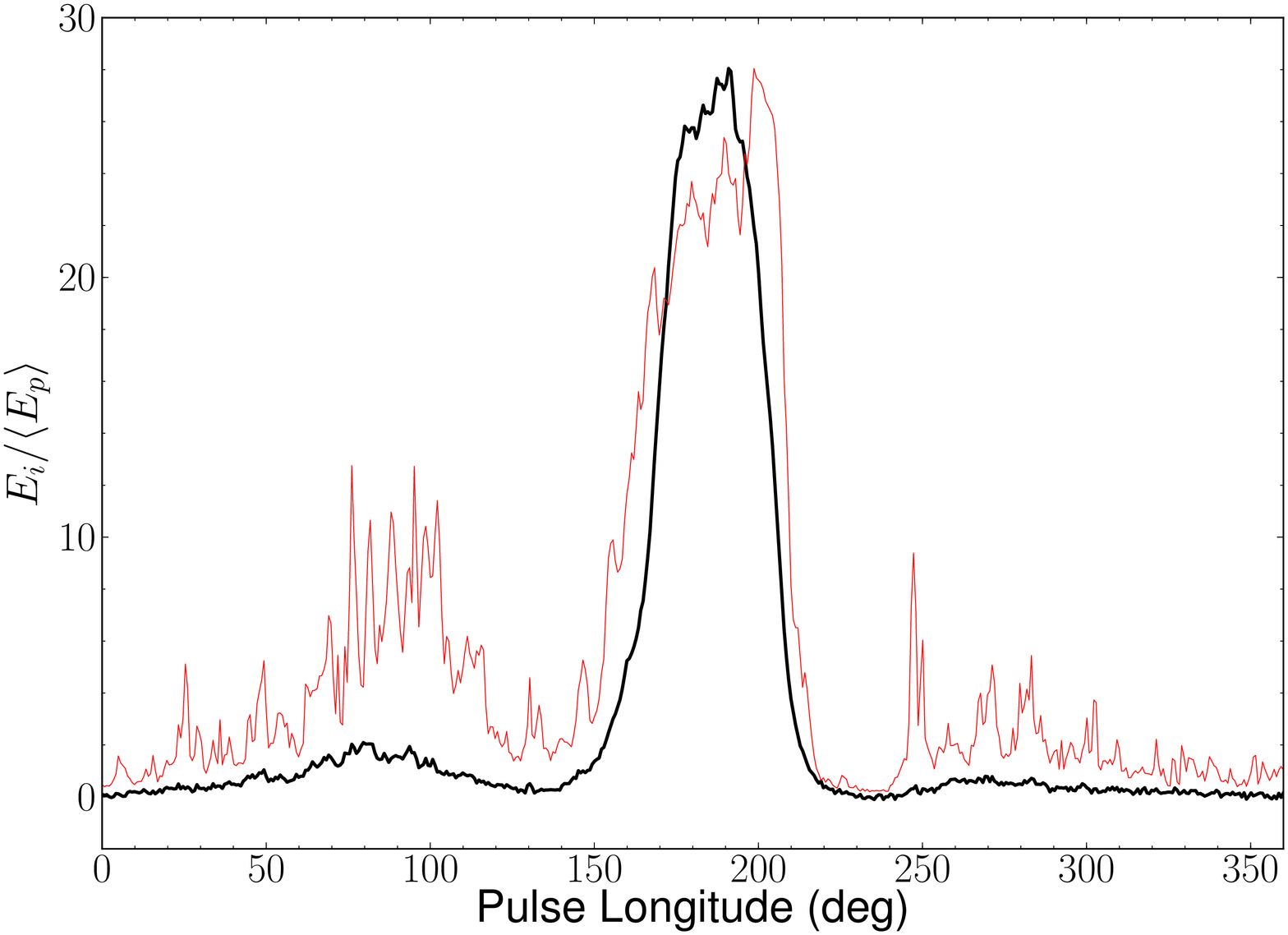}
   \includegraphics[trim = 1mm 1mm 1mm 3mm,clip,height=4.15cm,width=4.15cm,angle=270]{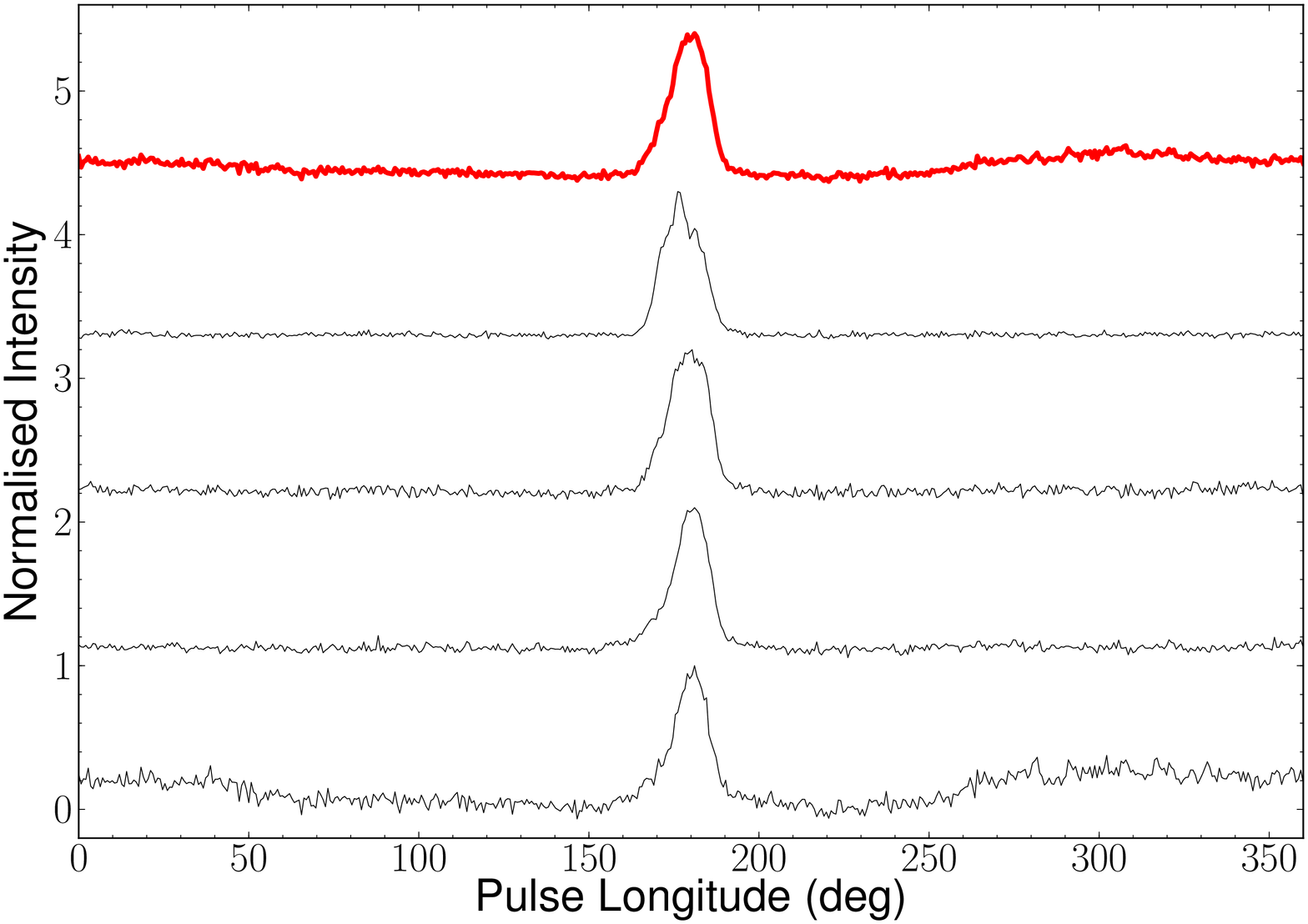}
  \includegraphics[trim = 1mm 1mm 1mm 3mm,clip,height=4.15cm,width=4.15cm,angle=270]{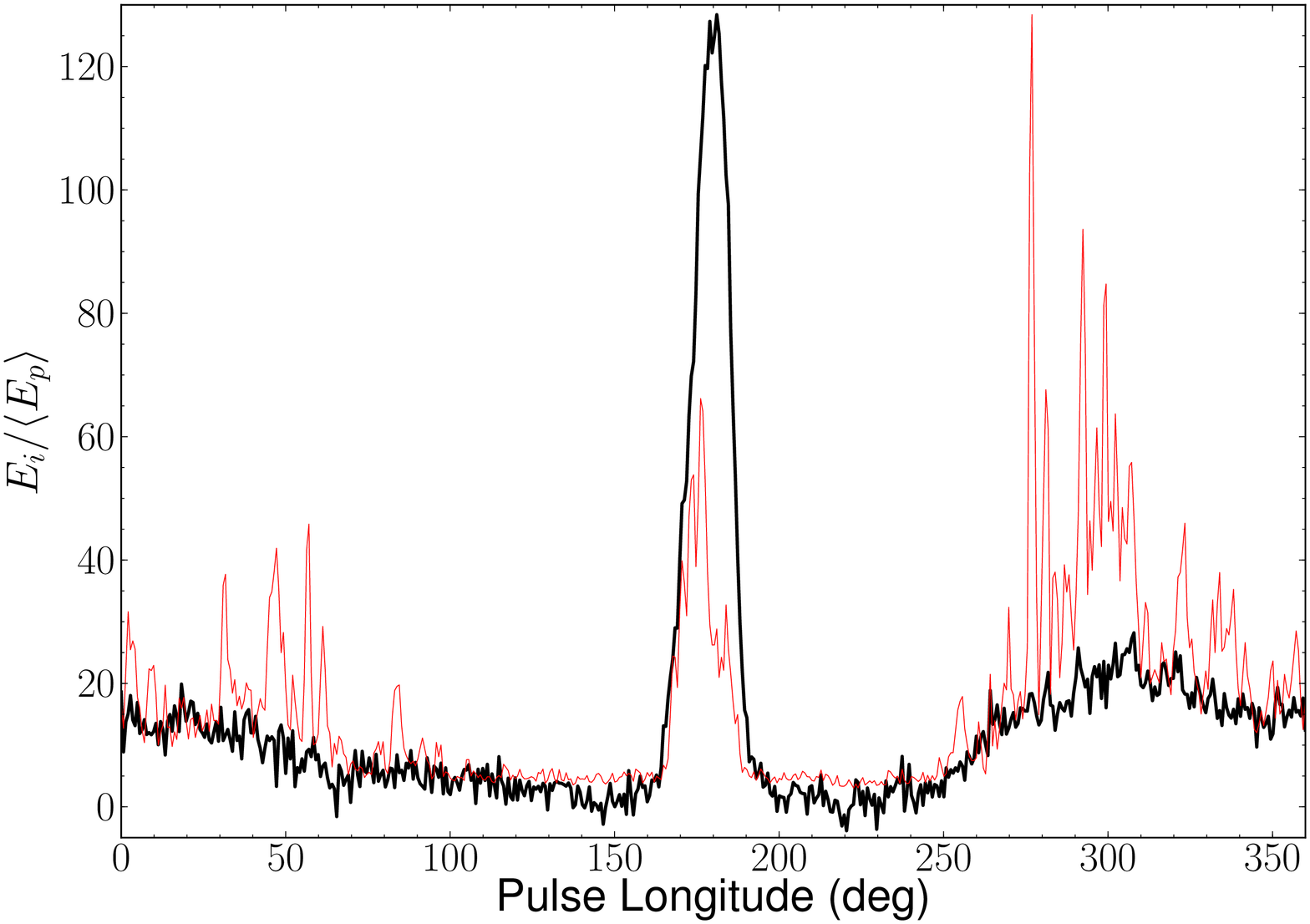}
\end{center}
\vspace{-6pt}
\caption{Pulse intensity fluctuation with respect to pulse longitude for the
  strong- (top panels) and weak-mode (bottom panels) pulses which exhibit
  $\geq6\sigma$ peak detections. \emph{Left:} the top thick line represents
  the average pulse profile, with successive lines (from top to bottom)
  denoting average profiles for pulses with $E>5\,\langle E\rangle$,
  $3-5\,\langle E\rangle$, $1.5-3\,\langle E\rangle$ and $0-1.5\,\langle
  E\rangle$, and for the strong mode contain 66, 34, 70 and 716 pulses,
  respectively. For the weak mode, these profiles contain 54, 157, 753 and
  2472 averaged pulses, respectively. Note that all the profiles are
  normalized by their peak intensity and are offset for clarity. \emph{Right:}
  scaled average profile (thick black line) and the brightest time sample for
  each pulse longitude bin (thin red line), compared with the average
  peak-energy of the profile $\langle E_{\mathrm{p}}\rangle$ (at a pulse
  longitude of $\sim180\degr$).}
\label{PEprofs}
\end{figure}

In several objects it has been shown that pulse profile variations can occur
as a function of pulse intensity (see e.g.
\citealt{gg01,elg+05,wwsr06}). Motivated by this possibility, we formed
integrated PE-separated profiles from the highest quality observations, for
increasing average on-pulse intensities. We averaged $6\sigma$ detected pulses
within energy ranges $E>5\,\langle E\rangle$, $3-5\,\langle E\rangle$,
$1.5-3\,\langle E\rangle$ and $0-1.5\,\langle E\rangle$ (where $\langle
E\rangle$ is the average pulse energy, over the whole on-pulse region, for a
given observation), as shown in Fig.~\ref{PEprofs}. The $6\sigma$ limit was
chosen to mitigate the effect of apparent nulls on the average pulse energies
and, subsequently, allow for the correct separation of pulses based on their
integrated energies.

In the strong-mode, we can clearly see that the average emission profile
becomes increasingly dominated by the MP component with increasing average
pulse intensity. By comparison, the weak-mode emission profiles become
increasingly more dominated by the PR and PC components with increasing
average intensity. This is also observed in both the 10- and 50-cm observing
bands. While the PC component is very prominent in the highest energy band,
and can often dominate over the MP, we note that the PE-separated plot for the
weak-mode does not reflect the typical properties of the source. That is, we
do not consider the average profile of the highest, weak-mode energy band to
be close to a stabilized profile. This is because there is only a low number
of pulses available for this analysis, coupled with the fact that only
$\sim60$~per~cent of the highest energy band pulses display emission in the PC
region. As such, we would expect the average PC component of the highest,
weak-mode energy band to be up to just over half the relative strength of the
average MP component, given a more stabilized profile through longer pulse
integration.

Using only the $6\sigma$ pulse detections for the mode-separated observations
again, we also determined the brightest pulse-energy sample for each pulse
longitude bin, with respect to the average peak-energy of the respective
profiles (see right panels of Fig.~\ref{PEprofs}). In the strong mode of
emission, we find that the brightest samples are preferentially distributed
among the MP and PR pulse-longitude regions. By comparison, we note that the
weak-mode emission is slightly more constrained. That is, the brightest
pulse-energy samples are preferentially distributed over a smaller proportion
of the separate emission regions during this emission mode, with the PC and MP
components dominating. Overall, it is clear that the pulsar emits across
almost the entire pulse longitude range.

\subsection{Fluctuation Spectra}\label{sec:FS}
\begin{figure*}
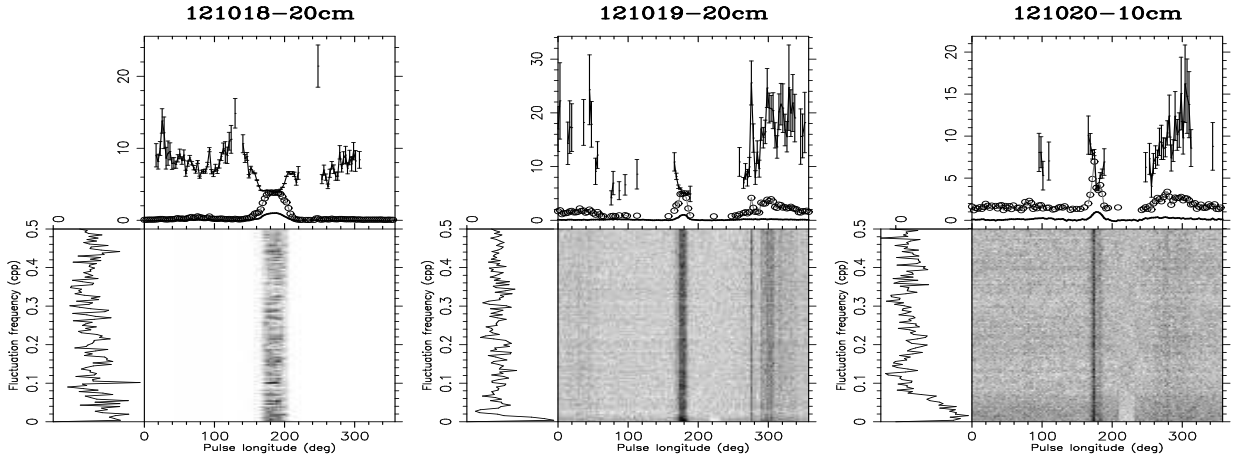

\begin{center}
  %% Normal trim dimensions are left,bottom,right,top
  % Trim dimensions _here_ are top,left,bottom,right
  \includegraphics[trim= 0mm -1.5mm 0mm 0mm, clip, height=5.5cm,width=6cm,angle=270]{121018-20cm_strong_lrfs.ps}%5-sig vals 
  \includegraphics[trim= 0mm -1.5mm 0mm 0mm, clip, height=5.5cm,width=6cm,angle=270]{121019-20cm_weak_lrfs.ps}%3-sig vals
  \includegraphics[trim= 0mm -1.5mm 0mm 0mm, clip, height=5.5cm,width=6cm,angle=270]{121020-10cm_lrfs.ps}%3-sig vals
\end{center}
\vspace{-5pt}
\caption{The modulation properties of PSR~J1107$-$5907, observed during the
  strong mode of 121018$-$20cm and the weak modes of 121019$-$20cm and
  121020$-$10cm (\emph{left to right respectively}). The integrated pulse
  profile (\emph{solid line}), longitude-resolved modulation index profile
  (\emph{solid line with error bars}) and longitude-resolved standard
  deviation profile (\emph{open circles}) are shown for each observation in
  the top panels. The bottom panels display the respective LRFS, with pulse
  longitude in degrees displayed on the horizontal axis. The side panels show
  the horizontal integration of the LRFS data, which represents the subpulse
  intensity modulation ($P/P_3$) of the pulse sequences.}
\label{FS}
\end{figure*}

While the source is highly variable, there does not appear to be any regular
periodicity in its intensity modulation. To test this hypothesis, we computed
longitude-resolved fluctuation spectra (LRFS; \citealt{bac70b}), as well as
two-dimensional fluctuation spectra (2DFS; \citealt{es02}) for several of the
strong- and weak-mode single-pulse observations. We calculated the LRFS by
taking Discrete Fourier Transforms (DFTs) along lines of constant pulse
longitude in the pulse stacks, over successive blocks of 256 pulses. The
resultant spectra were then averaged to provide a representation of the
typical modulation properties of the data, and have pulse longitude on the
horizontal axis and $P/P_3$ on the vertical axis (where $P_3$ is the subpulse
repetition period; see bottom panels of Fig.~\ref{FS}).

We also computed the longitude-resolved variance ($\sigma_i^2$) and modulation
index ($m_i=\sigma_i/\mu_i$) profiles\footnote{The uncertainty in $m_i$ is
  determined by bootstrapping $\sigma_i$. That is, additional random noise is
  incorporated into the data and, subsequently allows the variance in $m_i$ to
  be obtained.} (see top panels of Fig.~\ref{FS}) for the observations through
vertical integration of the LRFS ($\mu_i$ is the average intensity at a given
pulse longitude; see also \citealt{wes06} for more details). These parameters,
in combination with the LRFS, were used to infer the presence of any intensity
modulation and to determine whether it is random or periodic. To differentiate
between an intensity or phase modulation, DFTs were also performed on separate
pulse longitude regions within the LRFS to provide the 2DFS. The pulse
longitude range was again separated into three on-pulse regions, i.e. the PR,
MP and PC emission regions.

For both modes of emission, we find that the most prominent intensity
variation (i.e. highest modulation index) is typically associated with the PR
and PC emission components, where the emission is more sporadic. Appreciable
intensity modulation, similar to that seen in known pulsars, is also observed
across the shoulders of the MP component. This is shown by the distinctive
U-shapes in the modulation index profiles, which indicate that the dominant
intensity modulation in the MP component is located away from its central peak
(see also \citealt{wes06}). Note that the 50-cm data were too weak to perform
this analysis and were therefore excluded.

In the weak mode, the only significant LRFS feature can be observed in the
lowest frequency bin ($P_3\geq256~P$). However, we attribute this feature to
the baseline correction method used, given that the use of different running
mean lengths in the baseline normalization does not preserve this spectral
feature. In the strong mode, a couple of spectral features dominate over the
noise. The most significant of these is present at $P_3=9.6\pm0.1~P$. Further
investigation shows that this is a spurious signal which is only present in
one of the 256 pulse-long blocks of data. Therefore, we conclude that the
source only displays longitude-stationary non-periodic modulation.

\subsection{Polarization Properties}\label{sec:pol}
In a number of pulsars, emission moding and/or transient emission behaviour is
accompanied with changes in the source's polarization properties, such as the
presence of one or more orthogonal polarization modes (OPMs; see e.g.
\citealt{glr+92,krj+11,ksj+13}). With the above in mind, we sought to
characterize the polarization properties of the separate emission states of
PSR~J1107$-$5907, so that we might elucidate the emission variability of the
source. The results of this analysis are discussed below.

\subsubsection{Rotation Measure Considerations}\label{sec:RM}
The rotation measure (RM) of a source is the term used to quantify the degree
of Faraday rotation that its emission undergoes as it traverses through the
interstellar medium (ISM; e.g. \citealt{whl11} and references
therein). Faraday rotation, and hence the RM, can be quantified by measuring
the change in polarization position angle (PA) across a frequency band
(e.g. \citealt{njkk08}):
\begin{equation}
\Delta \mathrm{PA} = \frac{\mathrm{RM}\, c^2}{\nu^2}\,,
\label{eq:RM}
\end{equation}
where $c$ is the speed of light and $\nu$ is the frequency of the
electromagnetic waves. Following \cite{njkk08}, we measured the RM of
PSR~J1107$-$5907 in our polarization-calibrated observations using the
{\small\textsc{RMFIT}} package, which they developed as part of the
{\small\textsc{PSRCHIVE}} software suite \citep{hvm04}\footnote{See also
  http://psrchive.sourceforge.net/ for a detailed review.}. This package,
which uses a Bayesian likelihood test to find the best fitting RM to
equation.~(\ref{eq:RM}), obtains an $\mathrm{RM}=23 \pm 3$~rad~m$^{-2}$ for
our combined, time-integrated data. We note that this fit result serves as the
first published measurement of the RM of PSR~J1107$-$5907.

\subsubsection{Polarization Fluctuations}\label{sec:SP_pol}
\begin{figure*}
\begin{center}
  %% Normal trim dimensions are left,bottom,right,top
  % Trim dimensions _here_ are top,left,bottom,right
  \includegraphics[trim= 0mm 0mm 0mm 0mm, clip, height=8.3cm,width=7cm,angle=270]{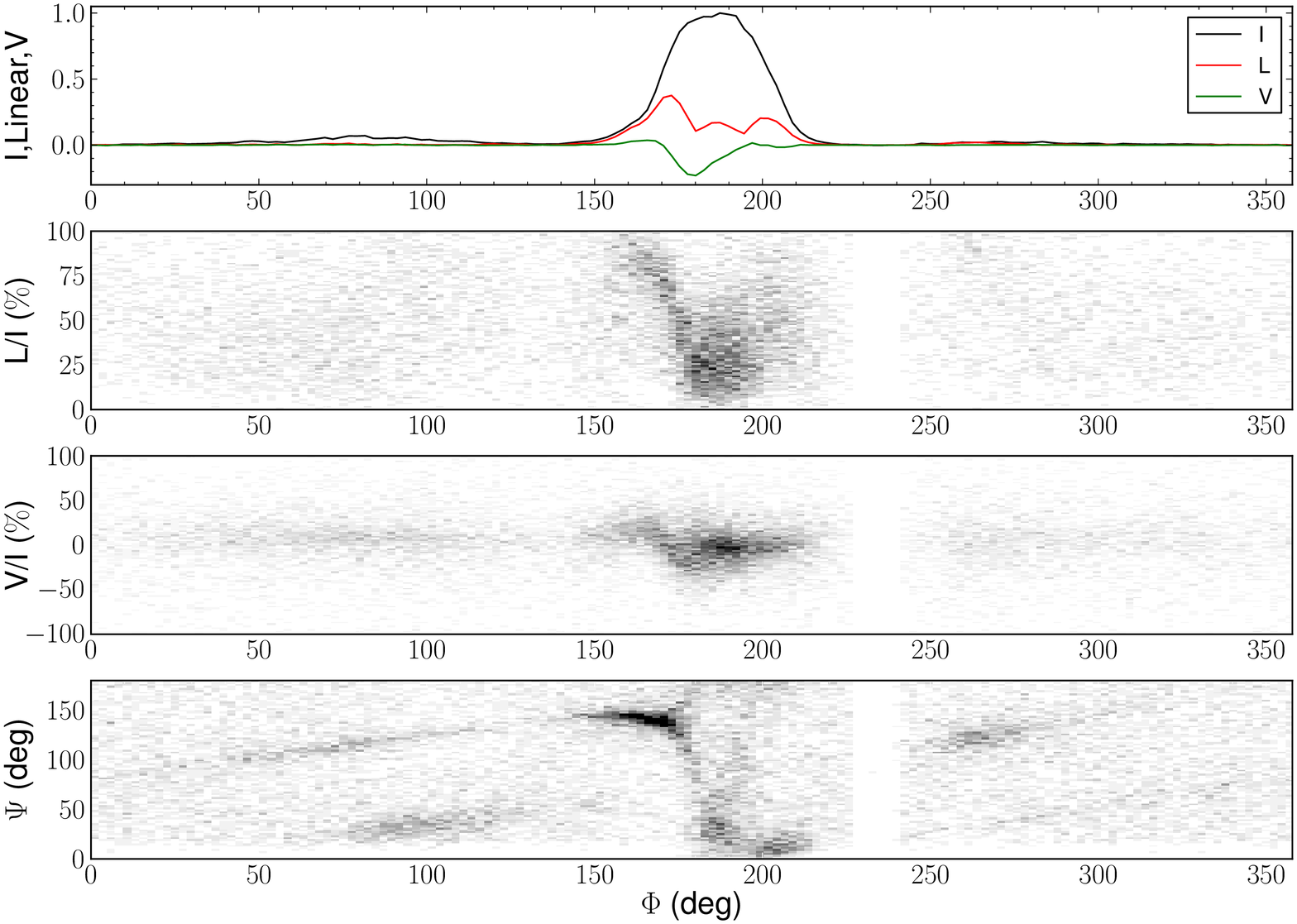} 
  \includegraphics[trim= 0mm 0mm 0mm 0mm, clip, height=8.3cm,width=7cm,angle=270]{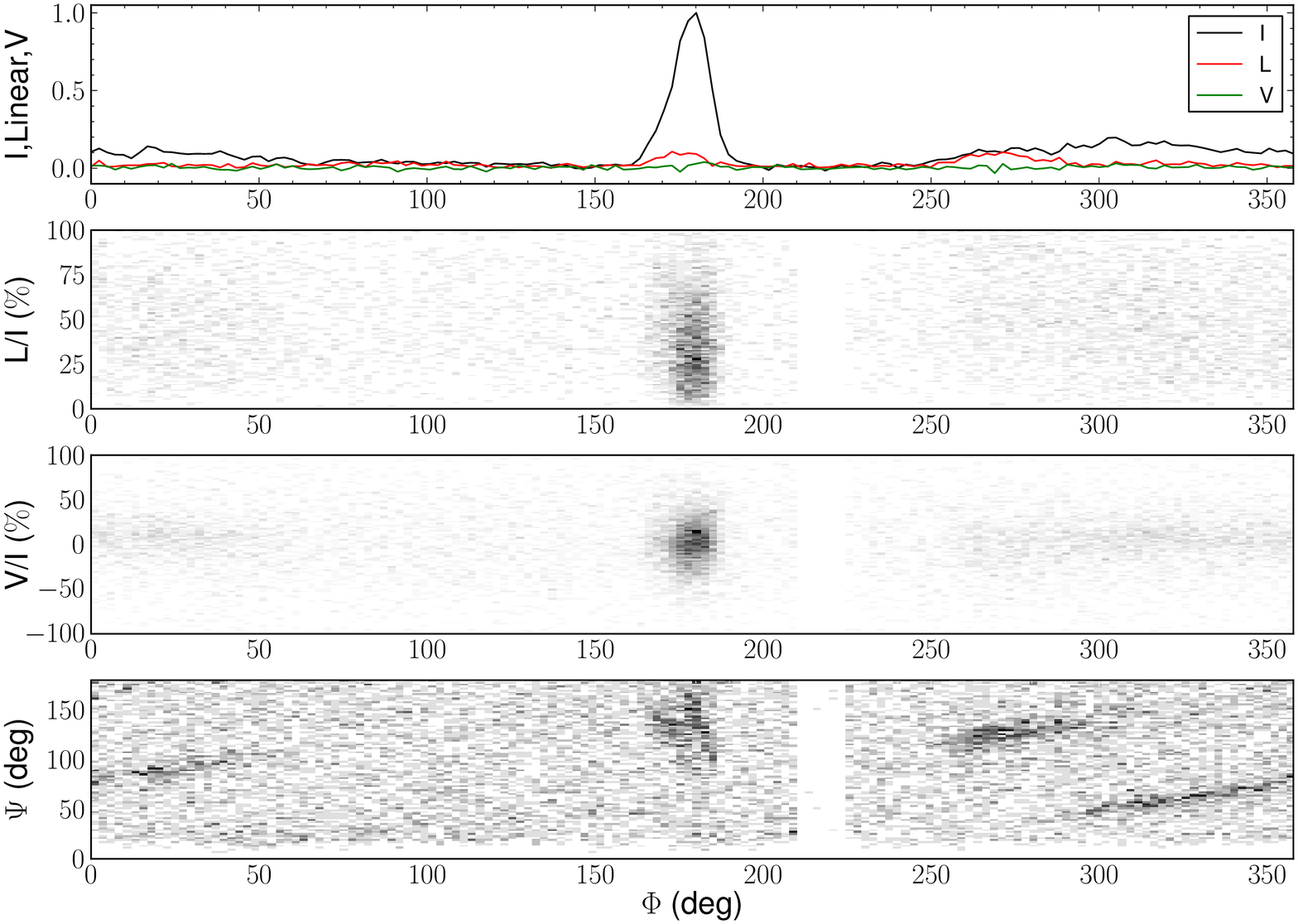}
\end{center}
\vspace{-7pt}
\caption{The single-pulse polarization properties of PSR~J1107$-$5907 in
  121018$-$20cm during its strong mode (\emph{left plot}) and its weak mode in
  121019$-$20cm (\emph{right plot}) respectively. For each of the plots, the
  average emission profile is displayed in the top panel. Single-pulse
  distributions of $L/I$, $V/I$ and PA with respect to pulse longitude are
  also shown (\emph{from top to bottom respectively}). Note the clear presence
  of the competing OPMs in the PR and PC regions for the strong and weak
  emission modes, respectively. The light vertical bands observed at fixed
  longitude throughout the data are an artefact of the baseline correction.}
\label{pol}
\end{figure*}

PSR~J1107$-$5907 exhibits very different polarization features during its
separate emission modes (see Fig.~\ref{pol}). That is, the strong-mode
emission features greater complexity, primarily in the MP component, and is
more highly polarized than that of the weak mode on average. In order to
ascertain whether there are any other polarization variations between the two
modes, we analysed the polarization calibrated single-pulse data, which are
capable of resolving short-time-scale fluctuations such as OPMs.

We measured the Stokes ($I$,$Q$,$U$,$V$) parameters and degree of linear
($L/I=\sqrt{Q^2+U^2}/I$; \citealt{br80}) and circular ($V/I$) polarization for
each pulse in our 20-cm data set. The PAs for these data were also measured,
so as to properly characterize the polarization properties of the
source. Given that random noise fluctuations can affect the reliability of
data samples, we only used pulses which contained $6\sigma$ detections. We
also restricted data samples to those with sufficient total intensities, i.e.
$\mathrm{SNR}(I)\geq3$, and required that the linear polarization components
have SNR values above a threshold of two for both modes of emission.  These
thresholds act to reduce the total number of data points available for further
analysis. However, they also act to significantly reduce noise contamination
in the distributions of $L/I$, $V/I$ and PAs, which enable the recovery of the
general polarization properties of the source, and facilitate further analysis
of the data.  Example polarization distributions from this analysis are shown
in Fig.~\ref{pol}.

From this analysis, we find that the pulsar emits radiation from at least two
competing polarization modes in both the strong and weak emission
states. During the strong emission state, these competing modes are shown in
the PR and PC regions. This is in contrast to the weak mode, where we only
observe two competing polarization modes in the PC region. Interestingly, we
also note the presence of non-OPM-like variations in PA in the central region
of the MP component (at $\phi\sim\ang{175}$ and $\sim\ang{195}$) during the
strong emission mode (c.f. PSR~B0329$+$54; \citealt{es04}). Overall, we see
that these variations are observed during the same observing runs and are only
found to coincide with the emission mode changes in the source.

Considering the above variations in PA, it is clear that the polarization
properties of the strong- and weak-mode pulses are quite
different. Furthermore, we note that the average polarization properties of
these pulses are dependent on the ratio of occurrence of the dominant OPMs
(see Fig.~\ref{PAfits}). This is supported by the analysis of the strong-mode
data from 121019$-$20cm, where only one competing PA-mode is observed over the
short sequence of pulses ($\sim 60$~s).

\subsubsection{Rotating-vector Model Fits}\label{sec:PAfits}
The magnetic inclination angle $\alpha$ and impact parameter $\beta$, of the
line of sight with the magnetic axis of a pulsar, can be used to define the
region where the observed radio emission is radiated from its
magnetosphere. As such, these parameters are central to the determination of
the emission geometry of a source. They can be constrained through fitting a
source's PA variation, as a function of pulse longitude, via the rotating
vector model (RVM; \citealt{rc69a,kom70}). This simple model takes advantage
of the close relationship between the PA and orientation of a pulsar's dipolar
magnetic field, to relate the rate of change of PA to the emission geometry of
a source \citep{kom70}:
\begin{equation}
  \tan(\Psi-\Psi_0) = \frac{\sin(\phi-\phi_0)\,\sin\alpha}{\sin\zeta\,\cos\alpha-\cos\zeta\,\cos(\phi-\phi_0)}\,,
\label{eq:PA}
\end{equation}
where $\Psi$ is the PA at a pulse longitude $\phi$ and $\zeta=\alpha + \beta$
is the inclination of the observer direction to the rotation axis ($\Psi_0$
refers to the PA at the longitude of the fiducial plane $\phi_0$). The
variation (or swing) in the linear PA, as the emission beam crosses our
line-of-sight (LOS), is normally expected to be monotonic and take the form of
an S-shaped curve \citep{rc69a}. However, non-RVM like features such as
90-degree jumps in PA (a.k.a. OPMs) can also be observed, as seen in
Fig.~\ref{pol} (see also, e.g., \citealt{scr+84,ls89}). While these features
increase the complexity of PA-swing fits, they are often consistent with the
RVM (see e.g. \citealt{lsg71,mth75,br80,scr+84}).

\begin{figure*}
\begin{center}
  %% Normal trim dimensions are left,bottom,right,top
  % Trim dimensions _here_ are top,left,bottom,right
  \includegraphics[trim= 2mm -1.5mm 2mm -3mm, clip, height=8.15cm,width=7.75cm,angle=270]{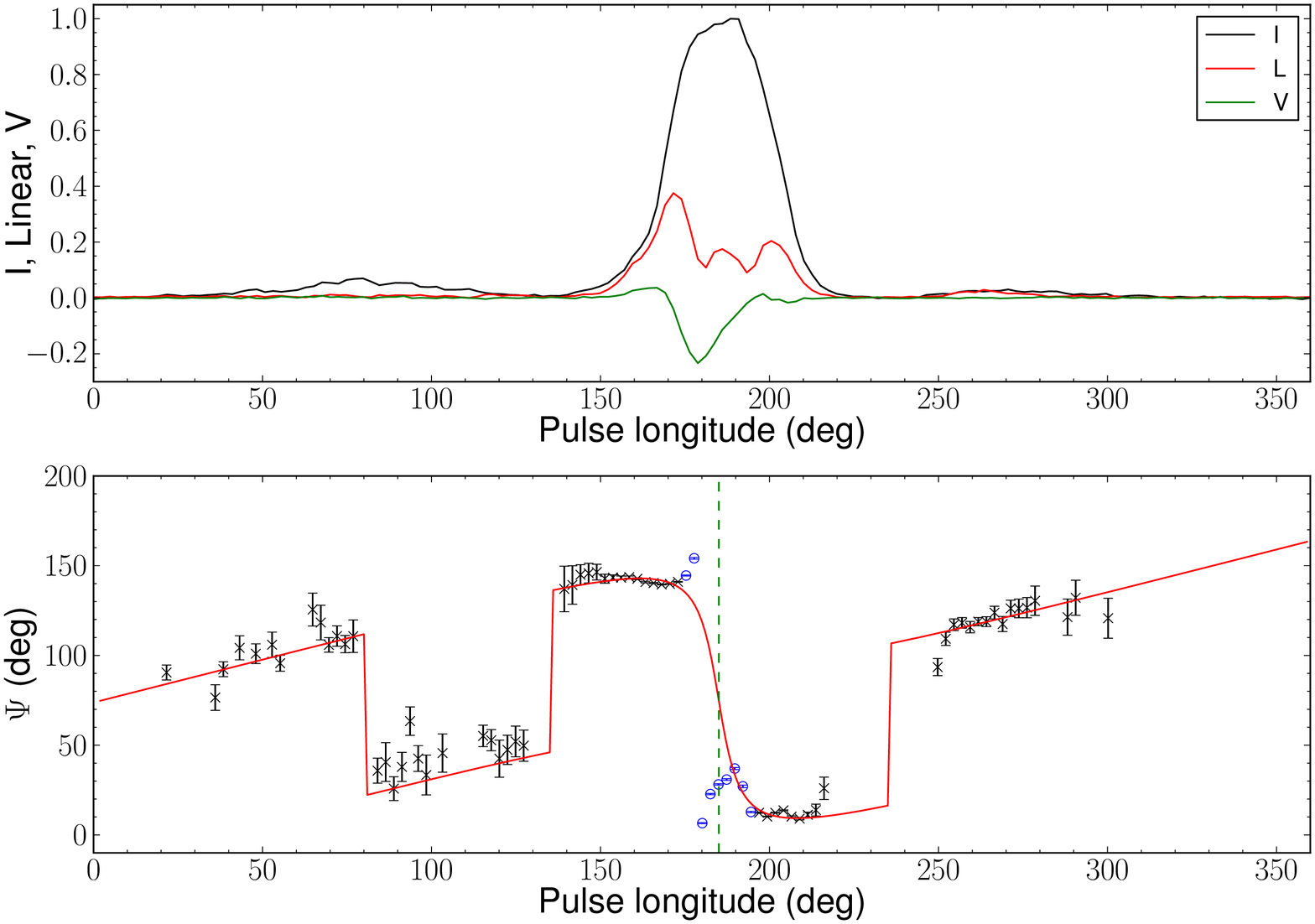} 
  \includegraphics[trim= 2mm -6.5mm 2mm 2mm, clip, height=8.15cm,width=7.75cm,angle=270]{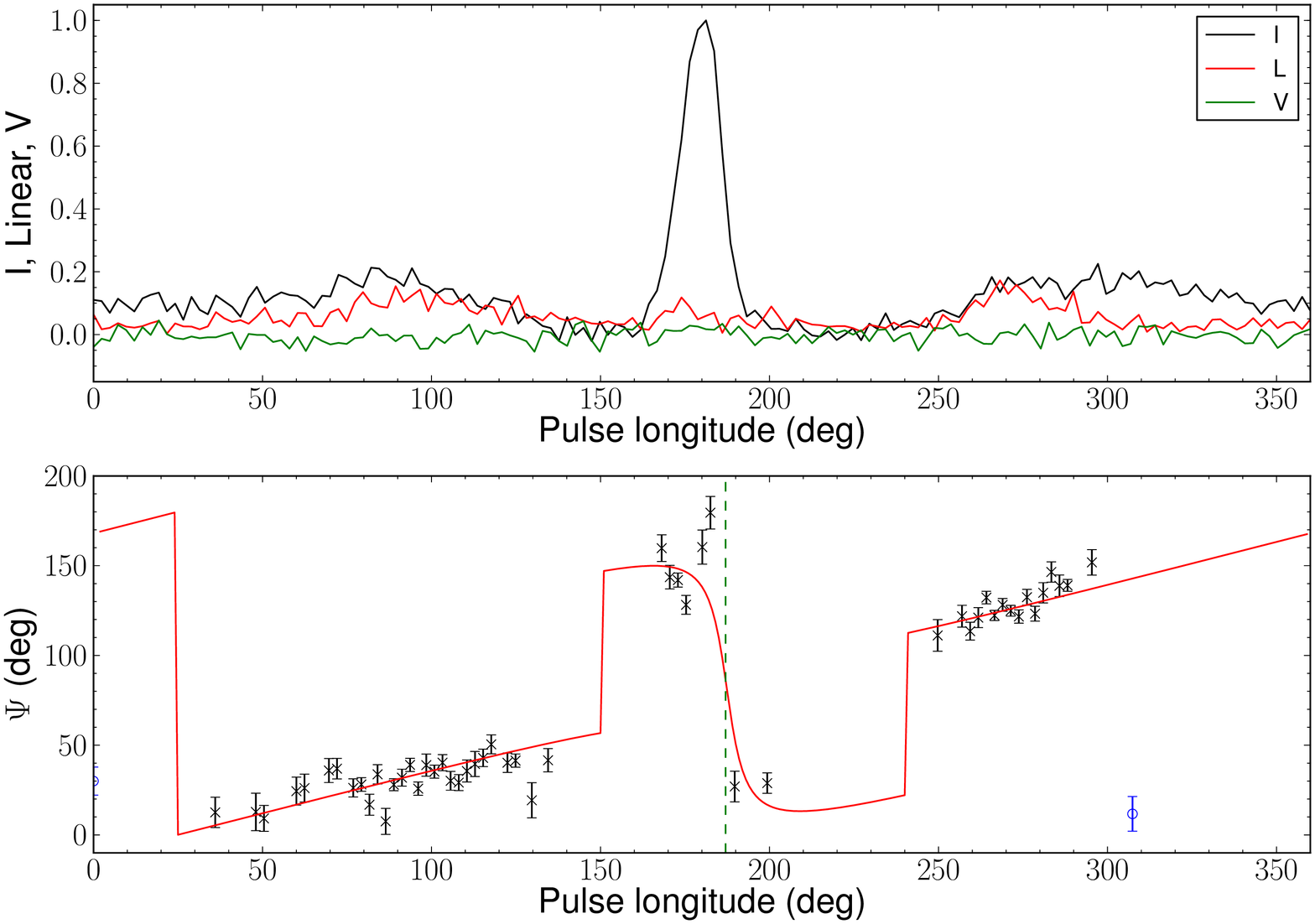}
  \includegraphics[trim= 7mm -2mm 0mm 0mm, clip, height=8.3cm,width=4cm,angle=270]{18oct12_20cm_strong_2sigMPhack_chi2.ps} 
  \includegraphics[trim= 7mm -2mm 0mm 0mm, clip, height=8.3cm,width=4cm,angle=270]{19oct12_20cm_2sigweak_2opms_chi2.ps}
\end{center}
\vspace{-7pt}
\caption{The best RVM fits to the integrated strong-mode data from
  121018$-$20cm (\emph{left}) and the integrated weak-mode data from
  121019$-$20cm (\emph{right}) respectively. \emph{Top panels:} the average
  emission properties of the mode-separated observations, showing the total
  intensity profile, as well as the linear and circular polarization
  profiles. \emph{Middle panels:} integrated PA data (crosses), with the
  best-fitting RVM (solid line) and location of the magnetic axis ($\phi_0$;
  dotted line) overlaid. PA data excluded from the fits are represented by the
  circles. Note the difference in location of the dominant PA distributions
  with respect to the separate modes of emission. \emph{Bottom panels:}
  $\chi^2$-fit surfaces for the RVM fits, showing the $\alpha$ and $\beta$ fit
  constraints. The contours display the $\alpha-\beta$ combinations which
  result in reduced $\chi^2$ values of $1\sigma$, $2\sigma$ and $3\sigma$
  within the nominal result.}
\label{PAfits}
\end{figure*}

Here, we use a $\chi^2$ minimization fitting method, based on
equation~(\ref{eq:PA}), to constrain the $\alpha$ and $\beta$ parameters for
PSR~J1107$-$5907 from our polarization-calibrated observations, by optimizing
$\phi_0$ and $\Psi_0$. For this analysis, we separated the time-integrated
observations by emission mode, after correcting for Faraday rotation, and
searched a grid of 200 by 200 possible $\alpha-\beta$ combinations for each
data set. In order to obtain the most significant results, we compromised
between the data quality and the number of fit points by only considering
strong- and weak-mode PA values above $2\sigma$ thresholds. We also do not
include strong-mode PA values from the pulse-longitude range
$\phi=\ang{175}-\ang{195}$ in the fits, due to the sharp decreases in linear
polarization and associated non-RVM consistent variations in PA (see
Fig.~\ref{pol}). The best results from this analysis are shown in
Fig.~\ref{PAfits}.

We find that the PA-swing of the pulsar emission is best fit with the RVM
using three orthogonal PA jumps (at $\phi=80\pm\ang{5}$, $135\pm\ang{5}$ and
$235\pm\ang{15}$) for the strong mode and two orthogonal PA jumps (at
$\phi=150\pm\ang{15}$ and $\ang{240}^{+10}_{-40}$) for the weak mode. Overall,
we note that the PA swings of the strong and weak modes are largely
consistent. The only noticeable difference between the two modes is the extra
OPM during the strong mode, which diverges from the predictions of the RVM.

Unfortunately, the range of fit parameters provided by $3\sigma$ limits from
the reduced $\chi^2$ plots does not result in very rigorous constraints. As
such, we are only able to place conservative limits on the emission geometry
of the source ($\alpha\gtrsim\ang{110}$ and
$\ang{-6.5}\lesssim\beta\lesssim\ang{0}$) from the mode-separated fits. We
also performed this analysis for combined strong- and weak-mode data from
121018$-$20cm and 121019$-$20cm separately. The results of this analysis are
consistent with the previous findings, but do not offer more stringent
constraints on the emission geometry of the source.

We note that the PA is seen over a wide range of pulse longitude from the
single pulses (refer to Fig.~\ref{pol}). However, we do not obtain better
constraints on $\alpha$ and $\beta$ from the RVM fits if we include the most
extreme PA values (e.g. from the $\Phi=\ang{310}-\ang{360}$ range). Rather, we
obtain equivalent constraints to those obtained from the average profile PA
curves (see Fig.~\ref{PAfits}).

Nevertheless, the above results are consistent with the source being a
near-aligned rotator. This interpretation is further supported by the source's
extremely broad emission profile ($\sim\ang{360}$) and advanced age
($\tau\sim447$~Myr), which are indications of magnetic alignment (see e.g.
\citealt{ran90,tm98,wj08,ycbb10,mgr11}).

\section{Timing Analysis}\label{sec:timing}
In a substantial sample of pulsars, clear correlations can be seen between
their pulse intensity/ shape and spin-down behaviour (see e.g.
\citealt{klo+06,lhk+10,ksj+13}). This leads us to suggest that similar changes
might occur in PSR~J1107$-$5907 if it is governed by the same magnetospheric
process(es) (see e.g. \citealt{lhk+10,lst12a}). To investigate such a relation
between pulse intensity and rotational stability, we calculated timing
residuals for PSR~J1107$-$5907 using our entire data set (see e.g.
\citealt{bh86,lk05} for details on this method). As the observations
displaying detectable emission contain a mixture of strong ($\sim10$~per~cent)
and weak ($\sim90$~per~cent) emission profiles, times-of-arrival (TOAs) were
calculated using two profile templates. These templates were formed from
analytic fits to the highest SNR strong- and weak-mode profiles using
{\small\textsc{PAAS}}\footnote{http://psrchive.sourceforge.net/changes/v5.0.shtml},
and were also aligned in time to remove any systematic offsets in measured
TOAs. The latter process, along with the computation of the timing residuals
(i.e. the difference between the observed and predicted TOAs) was carried out
using the {\small\textsc{TEMPO2}} package\footnote{An overview of this timing
  package is provided by \cite{hem06}. See also
  http://www.atnf.csiro.au/research/pulsar/tempo2/ for more details.}.

From this analysis, we find that the pulsar does not exhibit any significant
timing noise; i.e. the resultant timing residuals for our data set are white
(c.f. \citealt{hlk10}). We also obtain an average
$\dot{\nu}=-1.402\pm0.001\times10^{-16}$~s$^{-2}$, which is significantly
lower than that of pulsars with detected spin-down variation ($\sim10^{-15}$
to $10^{-13}$~s$^{-2}$; e.g. \citealt{lhk+10}). This finding is consistent
with the fits performed on the mode-separated TOAs, where we obtain spin
parameters which are consistent within the uncertainties of the fitting
procedure.

Following \cite{ysw+12}, we can approximate the minimum detectable spin-down
rate variation in PSR~J1107$-$5907 by
\begin{equation}
  |\Delta\dot{\nu}_{\mathrm{av}}| \gtrsim \frac{3\times\Delta\nu_{\mathrm{mod}}}{T}\,,
\end{equation}
where $|\Delta\dot{\nu}_{\mathrm{av}}|$ is the average change in $\dot{\nu}$
over the course of a period $T$, and $\Delta\nu_{\mathrm{mod}}$ is the
precision of our timing model. Assuming an ideal scenario where the pulsar:
(1) exhibits strong-mode bursts each of 24~min in length and a weak mode which
lasts 6-hr; (2) is observed continuously over the course of two strong- and
one weak-mode duration ($T\sim22480$~s); (3) can be optimistically timed to an
accuracy of $\sim10^{-9}$~Hz\footnote{The timing precision for approximately
  30~d of the best sampled TOAs in our data set is
  $\sim2\times10^{-7}$~Hz. However, with greater observing cadence this
  accuracy can be significantly improved.}, we would only expect to make a
$3\sigma$ detection for $|\Delta
\dot{\nu}_{\mathrm{av}}|/\dot{\nu}_{\mathrm{av}}\sim950$. This limit is
roughly 400 times greater than the largest spin-down variation currently seen
in any pulsar \citep{crc+12}. Furthermore, as the object exhibits its weak
emission mode for the majority of the time ($\sim96$~per~cent; see
$\S$~\ref{sec:DS}), the average spin-down rate of the object will be largely
determined by the $\dot{\nu}$ in this mode. As such, it is extremely unlikely
that the source would be able to experience such a large variation in
spin-down rate as the limit inferred above. With the above in mind, we surmise
that a variable $\dot{\nu}$ will be exceedingly difficult to detect in
PSR~J1107$-$5907.

\vspace{-3mm}
\section{Discussion}\label{sec:discuss}
\subsection{Giant-like Pulses?}\label{sec:GP}
From the analysis of the pulse stacks and flux density measurements, it has
been shown that PSR~J1107$-$5907 can emit very energetic, sporadic emission in
both its emission modes. The pulse energies of this particularly bright
emission are observed to far exceed the energy threshold of $10~\langle
E\rangle$, which is commonly used as an indication of giant pulse (GP)
detections (e.g. \citealt{cjd01}). However, we find a number of
dissimilarities between the bright emission of PSR~J1107$-$5907 and classical
GP emission. That is, we do not find evidence for a discernible break in the
distribution of peak flux densities (see e.g.
\citealt{ag72,lcu+95,ksv10}). Nor do we observe any confinement in the pulse
longitudes of the extremely bright pulses (see e.g.
\citealt{kbmo05}). Moreover, these pulses also exhibit large pulse widths
$\sim1.6-27.0$~ms (c.f. approximately $0.4$~ns~$-120~\mu$s for the Crab
pulsar; \citealt{kni07,he07, ksv10}) and, hence, large duty cycles that are
atypical of classical GPs ($10^{-2.2}\lesssim\delta\lesssim10^{-1.0}$, c.f.
$10^{-7.9}\lesssim\delta\lesssim10^{-2.4}$ for the Crab; \citealt{ksv10}).

As such, the brightest pulses attributed to PSR~J1107$-$5907 cannot be
considered classical GPs. Rather, we do find a strong relation with
PSR~B0656$+$14, which emits relatively wide pulses that can intermittently
exceed well above the GP threshold $10~\langle E \rangle$ (up to
$\sim116~\langle E\rangle$ in fact; \citealt{wwsr06}). As we will show in
$\S$~\ref{sec:RRAT} this suggests a connection with the RRAT population.

\vspace{-1mm}
\subsection{Detection Statistics}\label{sec:DS}
From the analysis of the longest observations, it is evident that the pulsar
is active in its strong emission state for only a small percentage of
time. During this mode, we see that the object preferentially emits bursts of
pulses, with typical apparent nulls of up to a few pulse periods. The emission
durations for the strong mode have been observed to be approximately
$1-24$~min in length, and appear to have a uniform distribution, with an
average duration $\langle T_{\mathrm{strong}}\rangle = 500\pm400$~s and highly
variable apparent $\mathrm{NF}\sim41-72$~per~cent. However, given the small
number of independent strong-mode detections (18), and number of observations
which do not completely cover strong-mode bursts (10), it is difficult to
accurately model these data.

Instead, we numerically estimated the best-fitting, average burst duration for
the observed detection rates. Here, we assumed that the pulsar exhibits two,
isolated strong-mode bursts over the course of a transit period at Parkes
($\sim11$~h~$27$~min), and that they appear uniformly distributed on a given
day, in-line with the observed detections. This results in an estimated
detection probability:
\begin{equation}
  P_{\mathrm{strong}} = \frac{N_{\mathrm{strong}}}{N_{\mathrm{transit}}}\,,
\end{equation}
where $N_{\mathrm{strong}}$ is the integer number of potential observations
containing strong emission ($\langle
T_{\mathrm{strong}}\rangle/T_{\mathrm{obs}}$), and $N_{\mathrm{transit}}$ is
the integer number of observations spanning the entire transit period
($T_{\mathrm{transit}}/T_{\mathrm{obs}}$). 

Using this method, we find that $\langle
T_{\mathrm{strong}}\rangle=740\pm20$~s results in the optimum number of
strong-mode detections. This corresponds to a total, average emission duration
of about 5900 pulses per Parkes transit period, and an inferred single-pulse
detection rate of $g_{\mathrm{strong}}\sim3.6$~per~cent in the 20-cm
band. Considering a typical observation length of $\sim30$~min for a RRAT
scan, we would then expect to obtain a strong-mode detection for every one in
12 observations in the 20-cm band. Further observation of the source at other
frequencies is required before any statistics can be inferred at other
observing wavelengths.

By contrast, detectable weak-mode pulses appear to be distributed uniformly
throughout observations, between apparent nulls of typically up to several
hundred pulse periods in length. As such, the UE during the apparent nulls in
this source will not be revealed until sufficient pulse integration is
performed. For our data, we obtain an increasing probability of weak-mode
detection with $T_{\mathrm{obs}}$ until $\sim80$~per~cent and above rates are
obtained for $T_{\mathrm{obs}}\gtrsim60$~min. This result is consistent with
the weak pulses assuming a log-normal intensity distribution (see
$\S$~\ref{sec:fits}), with increasing number of pulses contributing to more
significant detections. The low-level UE, or apparent null pulses, then
represents those pulses which are not individually detected in our data due to
intrinsic sensitivity thresholds (c.f. \citealt{eamn12} and references
therein).

However, the above does not provide the complete picture for the weak-mode
emission. If we only consider individually detectable weak-mode pulses
(i.e. $\gtrsim6\sigma$ detections), we obtain single-pulse detection rates of
$g_{\mathrm{weak}}\sim3$~per~cent in the 10- and 20-cm bands, and
$g_{\mathrm{weak}}\sim1.5$~per~cent in the 50-cm band from our data. This
corresponds to detectable weak-mode pulses being emitted at a rate of
approximately 1 weak pulse every 33 rotations, or $\sim430$~h$^{-1}$ in the
10- and 20-cm bands, and 1 weak pulse roughly every 67 rotations, or
$\sim240$~h$^{-1}$ in the 50-cm band. Therefore, this pulsar could be confused
as a RRAT-like source if it were only observed over short time-scales (see
also $\S$~\ref{sec:RRAT}).

\begin{figure}
\begin{center}
  %% Normal trim dimensions are left,bottom,right,top
  % Trim dimensions _here_ are top,left,bottom,right
  \includegraphics[trim= 0mm 0mm 0mm 0mm, clip, height=8.5cm,width=6.3cm,angle=270]{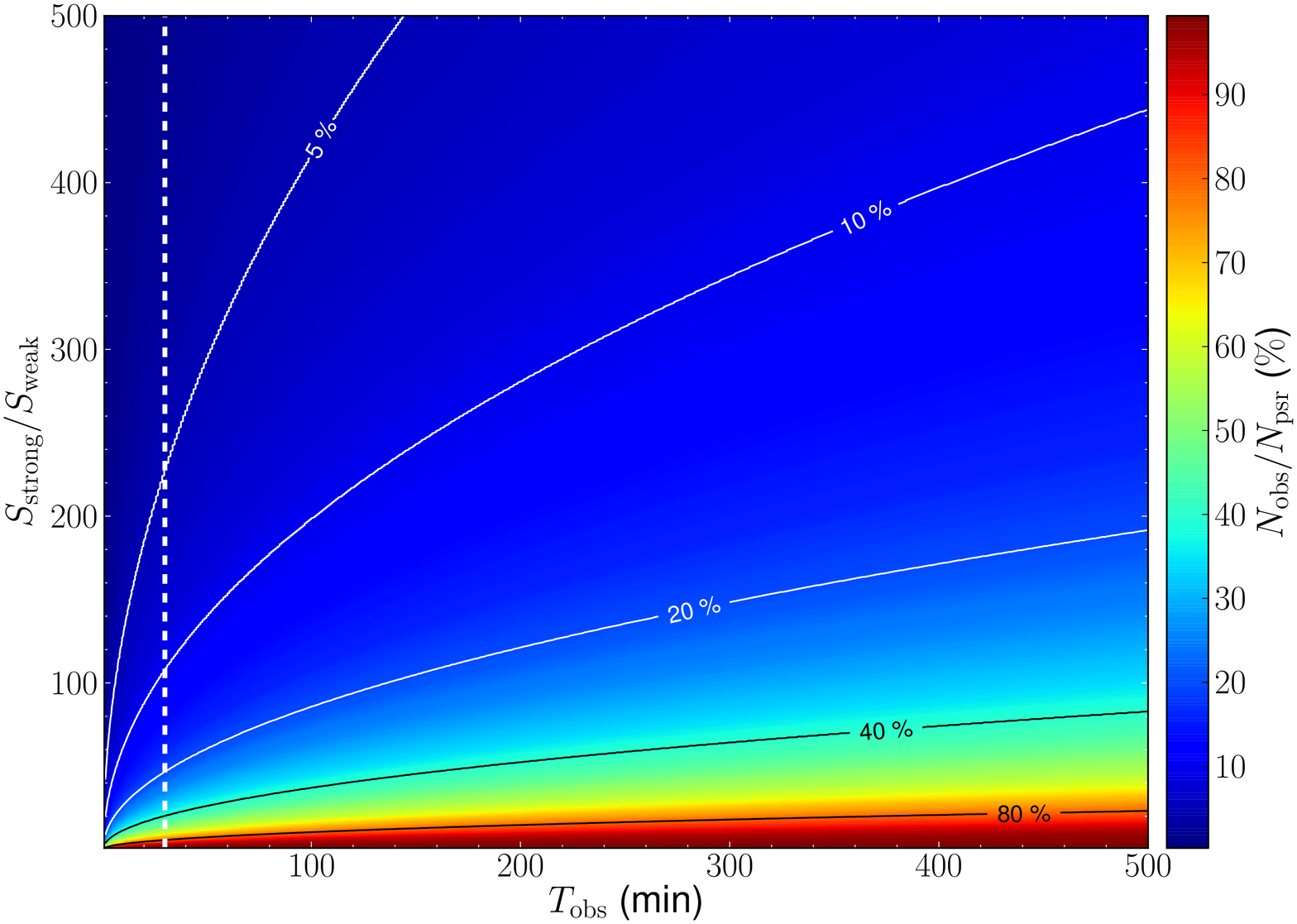} 
\end{center}
\vspace{-7pt}
\caption{Parkes Multibeam detection statistics for pulsars in the ATNF
  catalogue, assuming a $5\sigma$ detection threshold for a potential weak
  emission mode. The dotted line overlaid traces the detection statistics for
  an observation time of 30~min. Note that only $\sim11$~per~cent of sources
  would potentially be detected in a weak emission state assuming a flux
  density ratio of 100 between the strong and weak modes for this integration
  time. It is clear that the majority of sources would only be detected in a
  particularly weak emission state after substantial pulse integration, which
  is atypical of current intermittent studies.}
\label{detection}
\end{figure}

To further investigate the prospect of confusion between null emission and
very weak emission in the pulsar population, we sought to characterize the
number of sources which could be detected in a potential weak mode of
emission. In this context, we assume that all pulsars exhibit a strong and
weak mode of emission, with a flux density ratio of
$S_{\mathrm{strong}}/S_{\mathrm{weak}}$. Given that the Parkes telescope has
been the most successful pulsar survey instrument to date, we also assume that
observations are coordinated over a range of pulse integration time-scales,
with a telescope of the same size (i.e. 64~m), possessing system parameters
typical of the Parkes Multibeam receiver (see \citealt{mhb+13}). Furthermore,
we assume that this telescope can theoretically observe the entire known
pulsar population, which have defined period, equivalent width and flux
density parameters (refer to the ATNF catalogue; \citealt{mht+05}). From this
analysis, we find that only a very small fraction of the pulsar population (up
to $\sim11$~per~cent) could be detected by a 64-m telescope in a potential
weak state, assuming a flux density ratio of $\sim100$ between the strong and
weak modes and an observation time of 30-min (see Fig~\ref{detection}). Note
that the probability of detection becomes even lower for higher flux density
ratios, assuming the same integration time.

With the above in mind, it is clear that the detection statistics for many
sources are intrinsically linked to the length of observing runs.  This
provides strong motivation for increasing the typical length of observations
in sources which exhibit some form of moding behaviour and/or potential
nulling. It also suggests that the interpretation of nulls as true emission
cessation should coincide with the choice of observing system
(e.g. observation length and telescope) used, as previously mentioned by
several authors (see e.g. \citealt{kkl+11,bbj+11,lyn13}).

While there is yet no evidence for emission in the off-state of a substantial
number of intermittent pulsars
(e.g. \citealt{klo+06,crc+12,llm+12,ysw+12,gjk12}), we suggest that the
apparent null-states of many other known nulling sources should be studied for
the existence of low-level emission, given that they may in fact undergo
extreme mode changes without the need for the complete cessation of emission
(c.f. \citealt{elg+05,wmj07}). We further advance that even the deep
observations of several intermittent sources with existing telescopes (see
e.g. \citealt{klo+06,llm+12}) may not be sufficient to discover an extremely
low level of emission. This indicates that future telescopes such as the SKA
will be required to discern the true behaviour of moding and/or nulling
objects.

\vspace{-1mm}
\subsection{A RRAT Connection?}\label{sec:RRAT}
PSR~J1107$-$5907 shares a number of similarities with the RRAT population. As
shown above, the source exhibits a low single-pulse detection rate,
particularly in its weak emission state, which is consistent with the observed
detection rates of RRATs\footnote{See http://astro.phys.wvu.edu/rratalog/ for
  details on published RRAT data.}. Furthermore, the object displays extreme
brightness variations which are quantified by typical modulation indices that
are comparable to, or exceed those of RRAT-like sources (refer to
Table~\ref{tab:mod}, see also \citealt{wwsr06,ssw+09,wje11}). The peak
pseudo-luminosities of the pulsar, associated with the separate active
emission modes, are also consistent with the average associated with currently
known RRATs ($\langle L_{\mathrm{peak}}\rangle_{_{\mathrm{RRAT}}} =
5\pm8$~Jy~kpc$^2$; c.f. Table~\ref{tab:flux}).

\begin{table}
\caption{The average modulation properties of PSR~J1107$-$5907, with respect
  to emission region, for the observations chosen for spectra analysis.}
\begin{center}
\begin{tabular}{ c c c c c }
  \hline
  REF & Mode & $\langle m_{\mathrm{PR}}\rangle$ & $\langle m_{\mathrm{MP}}\rangle$ & $\langle m_{\mathrm{PC}}\rangle$ \\ \hline
  121018$-$20cm & Strong & $9\pm2$ & $6\pm2$ & $8\pm3$  \\
  121019$-$20cm & Strong$^a$ & $-$ & $-$ & $-$  \\\rule{0pt}{2.5ex}
  121018$-$20cm & Weak   & $11\pm4$ & $8\pm2$ & $15\pm6$ \\
  121019$-$20cm & Weak   & $16\pm3$ & $6\pm2$ & $16\pm5$ \\
  121020$-$10cm & Weak   & $7\pm1$ & $7\pm2$ & $10\pm3$ \\
  \hline
\end{tabular}
\end{center}
\hspace{6mm}$^{a}$Insufficient number of pulses to perform the analysis.
\label{tab:mod}
\end{table}

While the above similarities with conventional RRATs are interesting in their
own right, it is perhaps more interesting to compare the pulsar with objects
such as PSRs~J0941$-$39 and B0826$-$34. These pulsars are currently the only
sources to have been shown to exhibit `conventional' nulling and RRAT-like
behaviour \citep{bb10,bjb+12,eamn12}. PSR~J1107$-$5907 exhibits similar
behaviour to these pulsars, apart from the fact that there is discernible
emission present in its apparent nulls after sufficient pulse averaging
($\sim10^3$ pulses). The requirement to average over pulses to detect this UE
suggests that if the pulsar were placed farther away, then this source would
appear more similar to such RRAT-like objects (see e.g.
\citealt{wsrw06,kle+10}).

\begin{figure*}
\begin{center}
  %% Normal trim dimensions are left,bottom,right,top
  % Trim dimensions _here_ are top,left,bottom,right
  \includegraphics[trim= 0mm 0mm 0mm 0mm, clip, height=16cm,width=10cm,angle=270]{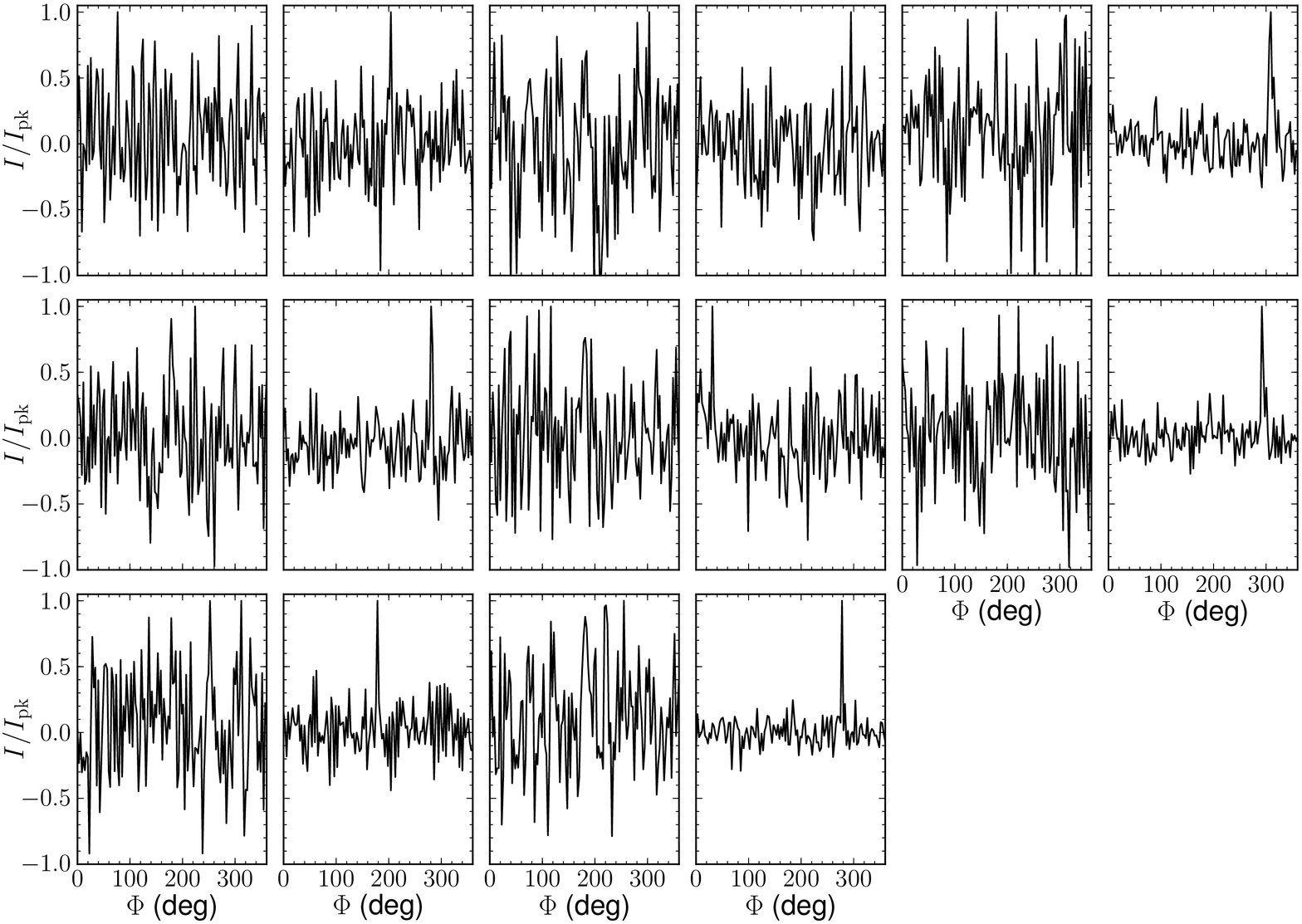} 
\end{center}
\vspace{-7pt}
\caption{Simulated emission profiles for the weak-mode emission of
  121019-20cm, which contain a factor of $\sim16$ increase in Gaussian noise
  compared with the observed data. From left to right, alternately, are the
  average profiles for successive 30-min integrations and the brightest
  single-pulse profiles for the corresponding pulse ranges. Note that the
  pulsar is rarely detected in the average profiles, if at all, while
  single-pulse detections are observed in the majority of the panels.}
\label{RRATprofs}
\end{figure*}

Indeed, when a factor of $\sim16$ increase in Gaussian noise is introduced to
the weak-mode pulses of PSR~J1107$-$5907 (i.e. mimicking a factor of $\sim4$
increase in Earth-pulsar distance), we find that the detection rates approach
close to zero for integrated groups of consecutive pulses ($\sim30$~min in
duration, c.f. typical RRAT scans). Whereas, the single-pulse detection rate
for the weak mode reduces to $\sim65$~h$^{-1}$ in the 20-cm band, when
introducing the same excess noise (see Fig.~\ref{RRATprofs}). This is in
contrast to the strong mode of emission, where the pulsar is detected in both
single pulses and in the average profiles, or not at all, depending on the
level of additional noise introduced. Therefore, the strong-mode of
PSR~J1107$-$5907 would likely represent the `pulsar-like' emission states of
PSRs~J0941$-$39 and B0826$-$34, and the weak mode would likely represent their
RRAT-like modes if the object were placed at a farther distance.

These findings provide additional support to the idea that RRATs are not a
distinct class of objects in the general pulsar population. Rather, they most
likely consist of a mixed population of modulated pulsars with extended PEDs
\citep{wsrw06} and extreme nulling pulsars \citep{bb10,kle+10}.

\vspace{-3mm}
\section{Conclusions}\label{sec:conc}
Our analysis of the emission behaviour of PSR~J1107$-$5907 has shown that the
source exhibits a very high degree of pulse-to-pulse variability, which is
comparable to that observed in RRAT-like objects. Remarkably, it has also been
shown that the flux density ratio of the average bright to weakest emission is
of the order of $\sim440:1$, which is considered to be a record in this
work. These attributes have led previous authors to suggest the presence of a
null mode of emission, in addition to the strong and weak modes observed in
our data. However, we have discovered low-level emission during the apparent
null phases of the longer weak modes, through integration of $\gtrsim10^2$
pulses which exhibit no discernible peaks (see $\S$~\ref{sec:mod}). This
emission resembles the weak-mode average profile and can, therefore, be
considered to be representative of emission from the lowest end of the PED for
the source. This indicates that the source most likely only exhibits two modes
of emission, with UE being present, during both the strong and weak modes of
emission. As such, we advance that the nulls observed in many intermittent
objects may just represent a transition to a particularly weak mode, rather
than the complete cessation of emission (see $\S$~\ref{sec:DS}).

We have also found that the source emits strong-mode pulses in isolated bursts
of $\sim200-6000$ pulses at a time, with the appearance of apparent nulls over
time-scales of up to a few pulses in between detections. While no discernible
emission was discovered through integration of these apparent nulls, we
advance that such emission should be unveiled after a sufficient number of
strong-mode pulses (i.e. $\gtrsim10^2$ pulses) containing apparent nulls are
integrated. We also infer a strong-mode detection probability of
$\sim8$~per~cent for observation scans of 30-min in duration.

During the weak mode of emission, we find that the source exhibits detectable
emission over time-scales of up to a few pulse periods at a time, with
apparent nulls typically lasting up to several hundred pulse periods. The
single-pulse detection probability for the source during this mode is found to
be $\sim3$~per~cent in the 10- and 20-cm bands, and $\sim1.5$~per~cent in the
50-cm band. This corresponds to single-pulse detection rates of
$\sim430$~h$^{-1}$ for observing wavelengths of 10 and 20~cm, and
$\sim240$~h$^{-1}$ at 50~cm. We also find that $\gtrsim1$~h integrations of
weak-mode pulses typically result in detections. This emphasizes the need for
long observing runs ($\gtrsim1$~h) when observing moding and/or transient
pulsars.

We also provide additional evidence for magnetic alignment in
PSR~J1107$-$5907. However, we stress that further polarization measurements of
this source are required to support this finding and fully map the
magnetospheric emission from the source.

Due to the low spin-down rate for this source, we did not detect a variable
$\dot{\nu}$. This follows from the findings of \cite{ysw+12}, who advance that
not all transient and/or moding objects will display detectable variations in
$\dot{\nu}$. As such, we suspect that only instruments such as the SKA will be
able to provide sufficient depth to the timing studies of such variable
pulsars.

Furthermore, we find that the pulsar can be reconciled with objects such as
PSRs~J0941$-$39 and B0826$-$34 if it were placed at a farther
distance. Coupled with the general similarities of the source with RRATs, this
further indicates that RRATs are most likely not part of a distinct class of
pulsars, as suggested by previous authors.

While it is not currently clear what the trigger mechanism for the behaviour
observed in PSR~J1107$-$5907 is, it is clear that future re-observation of
this source should help shed light on its behaviour. In particular,
high-energy studies of the source, during its strong mode, should prove
beneficial to constraining a potential driving mechanism, given the various
observationally verifiable predictions of each theory (see e.g.
\citealt{hhk+13}).

\vspace{-2mm}
\section{Acknowledgements}
We are grateful to S.~Oslowski, J.~Sarkissian, J.~Quick and D.~Yardley for
help in obtaining data which were used in this work. We would also like to
thank the members of the Parkes P786 observing programme for contributing
observations. Additional thanks go to R.~M.~Shannon and R.~Warmbier for
providing useful comments which have contributed to this research. NJY
acknowledges support from the National Research Foundation.

\bibliographystyle{mn2e} 
\bibliography{journals,psrrefs,njy_modrefs}
\end{document}